\documentclass[12pt]{article}

\usepackage{subfig}
\usepackage{times}
\usepackage{graphicx}
\usepackage{psfrag}
\usepackage{amsmath,amsthm}
\usepackage{amsfonts}
\usepackage{multirow}

\usepackage{mychicago}
\usepackage{genetics_manu_style}

\usepackage{xr}
\selectfont

\begin{document}

%\parskip=10pt

%\flushbottom 

\title{Predicting discovery rates of genomic features} 
\author{Simon Gravel\\Department of Human Genetics and G\'enome Qu\'ebec Innovation Centre\\ McGill University\\ Montr\'eal, QC, Canada, H3A 0G1\\on behalf of the NHLBI GO Exome Sequencing Project}

%\author{on behalf of the NHLBI GO Exome Sequencing Project\\Department of Human Genetics and Genome Quebec innovation Centre, \\ McGill University, Montreal, Quebec, Canada}
%\affil{1}{} } 

%

\renewcommand{\CorrespondingAddress}{Genome Quebec Innovation Centre\\
740, Dr. Penfield Avenue, Room 7206\\ 
Montr\'eal (Qu\'ebec) Canada\\
H3A 0G1\\

Phone: 514-398-7211\\
Fax: 514-398-1790\\
 \vfill}
\renewcommand{\RunningHead}{Predicting genome-wide discovery rates}
\renewcommand{\CorrespondingAuthor}{Simon Gravel}
\renewcommand{\KeyWords}{rare variants, capture--recapture,population genetics,linear programming,sequencing }

\date{}

\maketitle

\begin{abstract}

Successful sequencing experiments require judicious sample selection. However, this selection must often be performed on the basis of limited preliminary data. Predicting the statistical properties of the final sample based on preliminary data can be challenging, because numerous uncertain model assumptions may be involved.  Here, we ask whether we can predict ``omics" variation across many samples by sequencing only a fraction of them. In the infinite-genome limit, we find that a pilot study sequencing $5\%$ of a population is sufficient to predict the number of genetic variants in the entire population within $6\%$ of the correct value, using an estimator agnostic to demography, selection, or population structure. To reach similar accuracy in a finite genome with millions of polymorphisms, the pilot study would require about $15\%$ of the population.  We present computationally efficient jackknife and linear programming methods that exhibit substantially less bias than the state of the art when applied to simulated data and sub-sampled 1000 Genomes Project data. Extrapolating based on the NHLBI Exome Sequencing Project data, we predict that $7.2\%$ of sites in the capture region would be variable in a sample of $50,000$ African-Americans, and $8.8\%$ in a European sample of equal size. Finally, we show how the linear programming method can also predict discovery rates of various genomic features, such as the number of transcription factor binding sites across different cell types.  
\end{abstract}

%To estimate the number of individuals in an ecological population, a standard approach is to randomly sample subsets of the population, with replacement, on successive field trips. By keeping track of the frequency at  which captured individuals have been captured, it is possible to obtain estimates of total population sizes under a variety of scenarios. The simplest case, where all individuals are identical and have uniform and independent capture probability at each field trip, is a standard homework problem in introductory statistics courses. There is an extensive literature that addresses all the complications that may arise in real data: 
%\section{Significance statement}
%Rare genetic variants are central to many hypotheses about selection, demographic history, and the architecture of complex diseases. Because rare variants are more likely to be recent and deleterious, their number is used to define genetic burden in association studies and to infer recent population histories. 

%Estimating the number of variants in a population is therefore crucial for efficient sample design. We show how accurate estimates can be obtained based on sequence data from small subsets ($\simeq 5\%$) of the population. By contrast with previous methods, these strategies are robust to selection, demographic history, and population structure. We generate falsifiable predictions about the outcome of future sequencing experiments for diverse populations based on 1000 Genomes and NHLBI Exome Sequencing Project data.
\section{Introduction}
Predicting the genetic makeup of a large population sample based on a small subsample serves two distinct purposes. First, it can facilitate study design by providing the expected number of samples needed to achieve a given discovery goal, be it enough markers for a custom array design or enough rare variants to perform a well-powered burden test.  Second, such predictions serve as a useful test for our statistical and evolutionary hypotheses about the population. Because evolutionary experiments for long-lived organisms are extremely difficult, predictions about evolution are hard to falsify. By contrast, predictions about the outcome of sequencing experiments can be easily tested, thanks to the rapid advances in sequencing technology. This opportunity to test our models should be taken advantage of.  Here, we show that such predictions can be easily generated to high accuracy and in a way that is robust to many model assumptions such as mating patterns, selection, or population structure. 
%The robustness can accommodate intricate relatedness patterns among cases and controls, making the methods particularly appropriate for study design. The robustness also means that incorrect models can provide accurate predictions. Thus, any model-based prediction should improve upon the present nonparametric predictions to provide independent validation of model parameters.  

%predictions about the outcomes of additional sequencing often provides the only way through which  
%additional sequencing data often provides the most natural way to test proposed models can be teste.  
%such predictions are often the only way by which evolutionary hypotheses can be tested against subsequent measurements, i.e., the sequencing of additional samples. The rapid pace of sequencing technology advances provides a unique opportunity to do so, and this should be taken advantage of.   

We are interested in predicting the number of sites that are variable for some ``omic" feature across samples. Features may be of different types (SNPs, indels, binding sites, epigenetic markers, etc.), and samples may be cells, cell types, whole organisms, or even entire populations or species.  For definiteness, we will focus primarily on predicting the discovery rate of genetic variants (SNPs or indels) in a population. Because variant discovery is central to many large-scale sequencing efforts, many methods have been proposed to predict the number of variants discovered as a function of sample size in a given population. Some methods require explicit modeling of complex evolutionary scenarios, fitting parameters to existing data \cite{Durrett:2001it,Eberle:2000to,Gutenkunst:2009gs,Lukic:2011ie,Gravel:2011bg}. These approaches enable model testing, but they are complex and computationally intensive. The interpretation of model parameters can also be challenging \cite{Myers:2008fc}.  Ionita-Laza \emph{et al.} \cite{IonitaLaza:2009ik,IonitaLaza:2010jf} pointed out a similarity between the variant discovery problem and a well-studied species counting problems in ecology \cite{Pollock:1990tl}, and this led to the development of tractable heuristic approaches that rely on simple assumptions about underlying distributions of allele frequencies \cite{IonitaLaza:2009ik,IonitaLaza:2010jf,Gravel:2011bg} .  These methods are easy to use and often accurate, but the validity of the heuristic assumptions is uncertain, and departures from these models can lead to uncontrolled errors (see \cite{Link:2003wo} and the debate in \cite{Holzmann:2006cw}).  

%Detailed modeling of the SFS requires numerous assumptions and intense computational effort (see, e.g., \cite{{Lukic:2011ie}, {Gutenkunst:2009gs}} and references therein). 

%To address this, Ionita-Laza and colleagues \cite{IonitaLaza:2009ik, IonitaLaza:2010jf}, have therefore proposed a simplified predictive approach based on the assumption that the underlying allele frequency distribution is a beta-binomial distribution. In \cite{Gravel:2011bg}, we have proposed a linear estimator inspired by the  Burnham--Overton estimator  \cite{{Burnham:1979uv}}.  %Here, we show that jackknife predictions provide extremely accurate estimators for a diversity of human sample populations, and that a new linear programming approach provides more accurate estimates than was previously thought possible  [[\cite{Efron:1976tt, Link:2003wo}, as well as \cite{Holzmann:2006cw}]]

In this article, we build on the results of \cite{IonitaLaza:2009ik,IonitaLaza:2010jf,Gravel:2011bg} to propose improved estimators and quantify their uncertainties and biases. Even though fully nonparametric estimators were deemed impossible in the ecology problem (see \cite{Link:2003wo}, and Discussion), we obtain a nonparametric estimator based on linear programming (LP) that is asymptotically optimal in the infinite-genome limit, in the sense that the estimated confidence intervals contain precisely the values that are consistent with the data. These LP estimators are similar to estimators developed in the slightly different context of vocabulary size estimation (see \cite{Efron:1976tt}, and Discussion). Whereas parametric approaches were needed to get meaningful predictions beyond 10-fold extrapolation in the vocabulary problem, the nonparametric LP approach provide estimates of the number of genetic variants within $6\%$ of the correct value under 20-fold sample increases in a realistic genetic model in the infinite-genome limit, and within $35\%$ when $10^7$ polymorphisms are present in the entire sample.  We also present a jackknife-based estimator and provide strategies to estimate both the sampling uncertainty (via bootstrap) and bounds to the bias of the estimator. By applying the estimators to data generated by the 1000 Genomes Project (1000G) and the NHLBI Exome Sequencing Project (ESP), we find that both estimators compare favorably with the state of the art for computational efficiency, accuracy, and robustness to biases.

We provide examples of how these estimators can be used after preliminary data have been obtained to decide on the sample size required to achieve a given discovery goal, to estimate the impact of sample composition on projected study outcomes, and to predict the proportion of synonymous to non-synonymous sites as a function of sample size. Experimental design decisions require weighing many different factors, some of which must be estimated from incomplete information. Simple and robust estimates of the composition of the final sample should provide a useful tool for scientists seeking to obtain a clearer picture of the expected outcomes of different experimental strategies.

Finally, because nonparametric approaches do not depend on a specific evolutionary or biochemical model, they can be applied to a variety of genomic features. As an illustration, we apply the LP approach to predict the number of DNaseI footprints to be identified as a function of the number of cell types studied. Thus, the number of occupied transcription factor binding sites across all cell types in an organism can be estimated directly (and accurately) from a randomly selected sample of cell types. In addition to being a tool for study design, the discovery rate can answer fundamental biological questions, such as the total proportion of DNA that is bound by any or all transcription factors in any cell type.  

Software is available through the author's webpage.

\section{Methods}

Capture--recapture experiments use statistical inference to estimate population sizes without observing every individual. They use the overlap among random subsamples to estimate redundancy, and therefore of how much new information is to be found in unobserved samples. For example, the size of a rabbit population may be estimated by tagging $R_1$ randomly selected rabbits and counting the proportion $p$ of tagged rabbits in a subsequent random sample from the population. If rabbits and samplings are uniform and independent, the total population can be estimated as $R_1/p$. In practice, a number of complications may arise: sampling conditions may vary across field trips, rabbits can join or leave the population, and they can become afraid or fond of the capture device. As a result, an extensive literature on statistical methods accounts for these complications \cite{Pollock:1990tl}. A particularly challenging situation occurs when rabbits vary in their probability of capture. In this case, no amount of data can rule out the existence of a large number of very uncatchable rabbits. Based on this intuition, it has been argued that unbiased estimator for this problem required prior knowledge of the distribution of capture probability \cite{Holzmann:2006cw}. 

Ionita-Laza \emph{et al.} \cite{IonitaLaza:2009ik} pointed out that predicting the number of genetic variants that are present in a population is closely related to this rabbit-counting problem. In the analogy between the genetic and ecological cases, displayed in Table \ref{analogy}, rabbits to be counted are replaced by genetic variants; the capture of a rabbit is replaced by the variant identification in a sequenced (haploid) genome; and the probability of capturing a given rabbit on a given field trip is replaced by the population frequency of the variant. 

  \begin{table}
\centering
\caption{Some analogies between rabbit and genetic variant counting \label{analogy} }
\begin{tabular}{lr}
polymorphic loci & rabbits    \\	
\hline

sequenced chromosome	& sampling expedition  \\
nonreference genotype	& capture	\\					
allele frequency 		& rabbit catchability\\
rare variant			& rascally rabbit\\
%annotation			& species 	\\
\hline	
\end{tabular}
\end{table}

Whereas the ecological problem requires us to take into account the distribution of catchabilities among rabbits, the genetics problem requires us to  consider the distribution of allele frequencies among genomic loci. This distribution, $\Phi(f),$ depends on past mutation rates, demographic history, and selection;  and thus provides a natural testing ground for evolutionary models (see, e.g., \cite{Gutenkunst:2009gs,Lukic:2011ie} and references therein). The variant discovery rate therefore depends on many evolutionary parameters, but it is also limited by basic sampling statistics: in all models, the discovery rate per sample is expected to decrease as the sample size is increased. The goal of this article is to formalize this intuition and develop quantitative prediction methods. 

We consider a general model of ``omic" diversity which we will describe in terms of genotype diversity. A haploid genome has  $L$ independent loci. Each locus $i$ has a genotype $g_i$ that is ``reference" with probability $1-f_i$ and ``nonreference" with probability $f_i$. This nonreference allele frequency $f_i$ is drawn from an underlying frequency distribution $\Phi(f)$. To generate $n$ samples in this model, we would first draw each $\{f_i\}_{i=1,\ldots L}$ from  $\Phi(f).$ Then, for each locus $i$,  we would generate $n$ independent genotypes. We consider a variant ``discovered" if the nonreference allele has been discovered. An alternate definition, where a variant is discovered if both alleles have been observed in the sample, requires only minor modifications to what follows.

We do not know $\Phi(f)$ and would like to learn about it from the data. What we do observe is the sample \emph{site-frequency spectrum} (SFS), the histogram $\{\phi_n(j)\}_{j=1,\ldots,n}$ counting loci where exactly $j$ out of $n$ chromosomes have the nonreference allele in our sample. In the limit of infinite $L,$

\begin{equation}
\label{obs}
\phi_n(j)=\int_0^1 {n \choose j} f^j (1-f)^{n-j} \Phi(f) df.
\end{equation}

The SFS is a sufficient statistic for the unknown distribution $\Phi(f).$ We are now interested in predicting $V(N)$, the total number of variants discovered in a large sample of finite size $N$. 
Consider the number of undiscovered variants:

\begin{equation}
\label{basic}
V(N)-V(n)=\int_0^1\left((1-f)^n-(1-f)^N\right)\Phi(f)df.
\end{equation}
This is bounded below by 0, since the number of discovered variants must be positive. Because the rate of variant discovery per sample is expected to decrease with sample size, this quantity can also be bounded above. In the Supplement, we provide simple bounds that are based on generalizations of this argument and expressed as linear combinations of the $\phi_n(j)$; we refer to those as naive linear bounds. Even though they are mathematically interesting, we will see in Figure \ref{flower} that naive linear bounds do not provide the best practical bounds.  

%Since we are extrapolating to finite sample sizes, total counts are also bounded. Bounds that use linear combinations of the $\phi(d)$ are discussed in the Supplementary material. Even though these are mathematically interesting, we will see on Figure \ref{flower} that they do not provide the best practical bounds. 

\subsection{Linear programming}
Rather than think of our sample of size $n$ as drawn from an infinite population, imagine that it is drawn from a larger sample of size $N>n$, with allele frequency distribution $\Phi_N(i).$ In the limit of an infinite genome, the problem of finding values of $V(N)=\sum_{i\neq 0}\Phi_N(i)$ that are consistent with the observed $\phi_n(j)$ can be formulated as the linear program displayed in Table \ref{linprog}. This infinite-genome linear program always has a solution if the subsample was indeed generated by hypergeometric sampling from a distribution $\Phi_N(i)$. Since we have shown that $V(N)$ is bounded, the solution to the linear program is precisely the finite interval of values that are consistent with the data.  The existence of such an interval settles the question of whether estimates can be obtained without assumptions about the underlying frequency distribution \cite{Holzmann:2006cw}: \emph{point} estimators require assumptions about $\Phi_N(i)$, but interval estimators can be obtained using LP. If $N=\infty$, the intervals are semi-infinite. 
In practice, we can efficiently calculate tight bounds on $V(N)$ for $N$ in the thousands through the revised simplex method (see, e.g., \cite{Kasana2004}; here we use a version implemented in Mathematica). 
%We wish to find possible values of $V(N)$ that are consistent with the obser $\phi_n(j).$ This expected distribution of allele frequency in the sample is linearly related to the allele frequency distribution in the whole population $\Phi_N(i)$

\begin{table}
\caption{\label{linprog} The linear program formulation}
\begin{center}
\begin{tabular}{|c|c|}
\hline
Variables & $\Phi_N(i),~~~~~i\in\{0\ldots,N\}$\\
\hline
Objective function & $V(N)=\sum_{i\neq 0}\Phi_N(i)\equiv c_N \cdot \Phi_N$\\
\hline
Observational constraints &  
$\begin{array}{lcl}
\phi_n(j)&=&\sum_{i=1}^N  \frac{{i \choose j}{N-i \choose n-j} }{{N \choose n}} \Phi_N(i),\\
&\equiv& \mathbf{A_{N,n}}\cdot \mathbf{\Phi}_N.\end{array}$\\ 
\hline
Nonnegativity constraints & $\Phi_N(i)\geq 0.$ \\
\hline

%LP, $M=1000$, $\phi_M(0)$&78-154&96-106&99.9999-100.00011\\
%LP, $M=1000$, $\phi_M(0)=0$&78-117&96-103&99.9999-100.00008\\
%\hline
\end{tabular}
\end{center}
\end{table}

%$$ V(N)=\sum_{i\neq 0}\Phi_N(i)\equiv c_N \cdot \Phi_N$$
%over all underlying distributions $\Phi_N$ that are consistent with the observations $\phi_n(j)$:
%
%\begin{equation}
%\begin{split}
%\phi_n(j)&=\sum_{i=1}^N  {n \choose j} \left(\frac{i}{N}\right)^j \left(1-\frac{i}{N}\right)^{n-j} \Phi_N(i),\\
%&\equiv \mathbf{A_{N,n}}\cdot \mathbf{\Phi}_N.
%\end{split}
%\end{equation} 
%
%and to nonnegativity conditions 
%$$\Phi_N(i)>0.$$

%expected distribution of allele frequency in the sample is linearly related to the allele frequency distribution in the whole population $\Phi_N(i)$:

%We seek to obtain bounds on the objective function $ V(N)=\sum_{i\neq 0}\Phi_N(i)\equiv c_N \cdot \Phi_N$, given the set of constraints $\Phi_N(i)>0.$ 

An LP formulation of the capture--recapture problem was also used in a related problem of vocabulary estimation, where the sampling process is Poisson rather than hypergeometric \cite{Efron:1976tt}. By contrast to the Poisson case, where the unknown distribution of frequencies $\Phi$ is arbitrary, the underlying function $\Phi_N(i)$ in the genetics problem is usually drawn from a larger population of size $M$, and this imposes additional constraints on $\Phi_N(i)$ that can be incorporated into the linear program to improve accuracy:
\begin{equation}
\begin{split}
\mathbf{\phi_n}= \mathbf{A_{M,n}}\cdot \mathbf{\Phi_M}.
\end{split}
\end{equation}   
We wish to find an upper and a lower bound to the total number of variants. We must therefore solve two linear programs with the same constraints but opposite objective functions: $\pm c_N . \mathbf{A_{M,N}} \Phi_M $, where $c_N=\{0,1,1,\ldots,1\}.$  The resulting interval is the best possible estimator in the infinite-sites model for extrapolating from $n$ to $N$ in a population of size $M$, without using assumptions on the underlying model. In cases where we require a point estimator, we simply use the average of the upper and the lower bound. This is not entirely arbitrary---given the current constraints, the solutions at the constraint boundary have a frequency spectrum that reaches zero for some frequency (Figure \ref{minmax}) . We expect the correct value to lie in the interior of the interval.  

\subsection{Linear estimators}
The LP bounds are the best we can do without assumptions about $\Phi_N$. However, these may be computationally intensive for very large $N$.  Given the general success of the  Burnham--Overton (BO) jackknife estimators \cite{Burnham:1979uv}, it is worth asking whether similar estimators could be successful here. However, the BO assumptions that  $V(n)=\sum_{i=0}^p a_i \frac{1}{n^i}$ fail even for a panmictic, neutrally evolving constant-size population (i.e., the Standard Neutral Model, where $V(n)\simeq \log(n)$). % (see supplementary figure X). 

In \cite{Gravel:2011bg}, we proposed an expansion of the form  $V(N)-V(n)=\sum_i a_i \left(H(N)-H(n)\right)^i$, with $H(n)=\sum_{i=1}^{n-1}1/i,$ the $(n-1)$st harmonic number. A simpler and more principled expansion is $V(n)=\sum_{i=0}^p b_i H^i(n).$ We show in the Appendix that both expansions yield the same jackknife estimates, but the latter is more tractable. Even though more general expansions could be considered, this particular expansion is practical because a) it provides exact results at linear order in the Standard Neutral Model, b) it allows the modeling of a diversity of functions that increase slowly but do not converge, and c) it performs well in simulations (Figure \ref{SQRT}). We refer to the resulting estimate as the harmonic jackknife.

\subsection{Finite genome}
Two complications can arise as a consequence of the finiteness of the genome. First, the infinite genome approximation underlying the Standard Neutral Model expression $V(n)\sim \log(n)$ that serves as a starting point for the jackknife expansion may not hold: for a large enough sample size, we will run out of sites. The BO estimator might eventually become a better choice. The LP approach would not be sensitive to this problem, as it does not rely on the Standard Neutral Model. 

The second complication introduced by a finite genome is that the observed site-frequency spectrum is now a random variable, as there are a finite number of observations per frequency bin. For the jackknife estimator, this may result in large, uncontrolled inaccuracies, especially if high-order estimators are used. The infinite-sites LP problem, by contrast, is likely to be infeasible in the presence of noise.  Under the Random Poisson Field approximation, one may attempt to maximize the likelihood 
$$L[\Phi]=\prod_{i=1}^n P((A\Phi)_i, \phi(i)),$$
 under the constraint $\Phi \geq0,$ where $P(\mu,x)$ is the Poisson distribution with mean $\mu$. The maximizing $\Phi$ may or may not be unique (so that we may have either a point or an interval estimator). Unfortunately, because the optimizing problem is now nonlinear, the general optimization problem is intractable in its exact form. 

To take advantage of the LP formalism, we may wish to relax some of the constraints imposed as equalities in the infinite-$L$ limit, in such a way that realizable vectors exist and the LP problem can be solved. One approach is  to turn equality constraints into range constraints \cite{Efron:1976tt}, with width informed by the expected fluctuation sizes in each bin. However, a more efficient option is to coarsen the least informative bins. Since most of the unobserved variants are rare, we do not care for the precise frequency of the common variants. We use a bin-merging strategy, collapsing bins containing common variants into a smaller set of coarser bins. This has the added benefit of reducing the number of constraints, making the problem numerically more tractable. We use a simple scheme in which we keep the $p$ lowest-frequency bins intact, then merge the next two bins, then the following four bins, and so on,  increasing the bin size exponentially until all bins have been taken into account.   We then choose $p$ as high as possible without making the LP problem infeasible. Fortunately,  Figure \ref{lpbybin} shows that it is not necessary to use a large number of bins to obtain tight bounds.

This procedure will result in a predicted range for the number of discovered polymorphisms. This range accounts for uncertainties about the underlying distribution, but not for sampling uncertainty. To account for sampling uncertainty, we can bootstrap the data, each bootstrap iteration providing a confidence interval. We can then define confidence intervals by using $95\%$ confidence intervals on both the upper and lower bounds. Such confidence intervals on bounds are expected to be more conservative than confidence intervals on point estimates.

\subsection{Multiple populations}
The strategies described above do not require random mating assumptions. They can therefore predict the number of variants in samples drawn from multiple populations if subsamples from the subpopulations are available.  The LP approach can be generalized to bound any linear function of the joint SFS, including the number of variants private or share across samples. However, the number of variables grows rapidly with the number of populations, and such a linear program would require careful optimization. We use a simpler strategy and form a subsample with the appropriate ancestry proportions, and extrapolate using the single-population strategies outlined above. In Figure \ref{admix}, we show extrapolations based on 100 African and 100 European haplotypes, using 1000 Genomes YRI and CEU populations, as well as results based on equal-sized samples using a known simulated demographic history. As expected, we find that discovery rates are higher in mixed populations, and the mixing proportion that maximizes discovery depends on the total sample size.

\subsection{Alternate approaches}

We compare the results of the methods presented above to three different strategies: a) the parametric model of Ionita-Laza \emph{et al.} \cite{IonitaLaza:2009ik}, which supposes that the allele frequency distribution can be modeled as a Beta distribution with parameters fitted to the observed distribution of allele frequencies; b) the standard  Burnham--Overton estimator of order 3, which supposes that the proportion of missed variants at sample size $N$ can be expanded as a third-order polynomial in $\frac{1}{N};$  and c) a fully model-based approach, using $\partial$a$\partial$i \cite{Gutenkunst:2009gs} to fit a three-parameter, one-population demographic model to the observed SFS. The model involved  two periods of constant population size, $N_1$ and $N_2$, and instantaneous change between the two values at time $t$.

\section{Results}
\subsection{Simulations}
\begin{figure}
\scalebox{.6}{\includegraphics{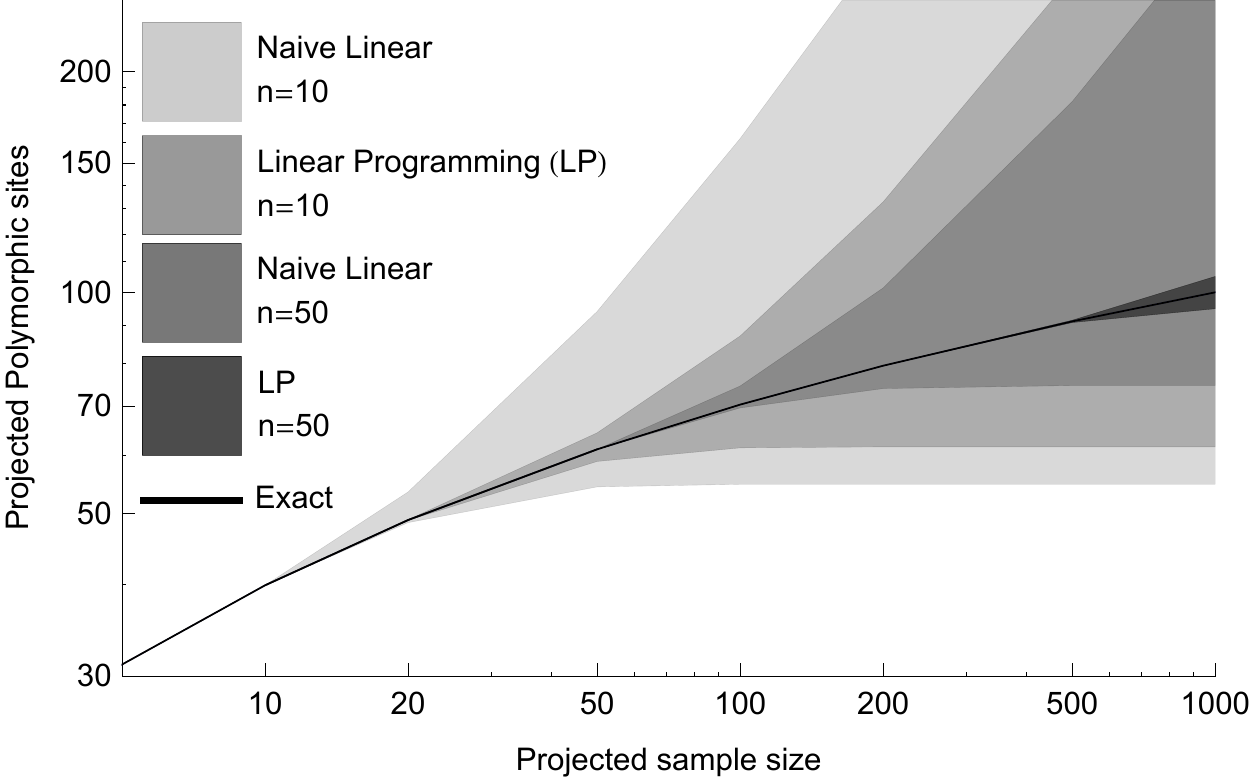}}

\caption{\label{flower} Bounds on the number of polymorphic sites to be discovered, based on discoveries in samples of size $n=10$ and $n=50$ from model \eqref{simu} using second-order naive linear and linear programming (LP) bounds. The shaded areas correspond to strict upper and lower bounds in the infinite-sites model and do not rely on any assumption about distribution \eqref{simu}. Note that LP provides dramatically tighter confidence intervals for both sample sizes, and that order-of-magnitude extrapolations can be performed for LP and $n=50$ with high accuracy.}
\end{figure}

To study the predictive power of different methods in the infinite-sites limit, we generated expected frequency spectra in a population of $M=1000$ individuals, with 

\begin{equation}
\label{simu}
\Phi_{1000}(i)\propto \frac{1}{i+0.1},
\end{equation}
and for subsamples of size $n\in\{10,20,50\}$. Extrapolations were attempted to $N\in\{20,50,100,200,500,1000\}.$   Figure \ref{flower}  presents extrapolations based on samples of size 10 and 50 using naive linear and LP bounds, and  Table \ref{tabex} shows confidence intervals for extrapolations to $N=200$ using two different naive linear bounds: LP, and LP using the $M>N$ strategy. To facilitate comparison, the predicted number of polymorphisms is expressed as a percentage of the variants in the population of 1000 individuals. Because these simulations closely follow the harmonic jackknife assumptions, harmonic jackknife estimates are essentially perfect, but this is not representative. Harmonic and  Burnham--Overton jackknife estimates with different underlying distributions are presented in Figure \ref{SQRT} and in the 1000 Genomes example below. 

%\begin{figure}
%\scalebox{.5}{\includegraphics{./figures/quadrat_best}}
%\scalebox{.5}{\includegraphics{./figures/cubic}}
%\scalebox{.5}{\includegraphics{./figures/lin_prog_vanilla}}
%\scalebox{.5}{\includegraphics{./figures/best_1000_LP}}
%\caption{Extrapolation bounds for an underlying model with 1000 samples and $\Phi_{1000}(i)\propto \frac{1}{i+0.1}$, based on subsamples of $10, 20$, and $50$ samples. From top to bottom: The best linear upper and lower bounds with $d=2$,... }
%\end{figure}

%\begin{tabular}{|c|c|c|c|}
%\hline
%N=100  & \multicolumn{3}{c|}{Sample Size (n)}\\
%\hline
%Method &10&20&50\\
 %\hline
%Jackknife, $d=2$&55-162&63-162&69.6-103\\
%Jackknife, $d=2,3$&55-100&63-78&69.6-70.7\\
%LP, $M=N$&58-91&68-72&70.2981\\
%LP, $N=1000$&61-87&70.0-70.6&70.2981\\
%\hline
%\end{tabular}

%\begin{tabular}{|c|c|c|c|}
%\hline
%N=200  & \multicolumn{3}{c|}{Sample Size (n)}\\
%\hline
%Method &10&20&50\\
% \hline
%Jackknife, $d=2$&55-298&63-171&74-101\\
%Jackknife, $d=2,3$&55-161&63-111&74-84\\
%LP, $M=N$&60-137&73-87&79.416-79.439\\
%LP, $M=1000$&62-133&76-84&79.4285-79.4286\\
%\hline
%\end{tabular}

\begin{center}

\begin{table}
\caption{\label{tabex} $100\%$ confidence intervals for extrapolating the number of polymorphic sites discovered in 200 chromosomes, based on samples of size $n=10, 20, 50$ and four different approaches, in the infinite-sites limit. The intervals are expressed as percentages of the correct value from the model given in Equation \eqref{simu}}
\begin{center}
\begin{tabular}{c|c|c|c}

%\hline
N=200  & \multicolumn{3}{c}{Sample Size (n)}\\
\hline
Method &10&20&50\\
 \hline
Naive linear, $d=2$&69-375&79-215&93-128\\
Naive linear, $d=2,3$&69-203&79-139&93-106\\
LP, $M=N$&76-173&92-109&99.98-100.01\\
LP, $M=1000$&78-167&96-106&99.9999-100.0001\\
\hline
%LP, $M=1000$, $\phi_M(0)$&78-154&96-106&99.9999-100.00011\\
%LP, $M=1000$, $\phi_M(0)=0$&78-117&96-103&99.9999-100.00008\\
%\hline
\end{tabular}
\end{center}
\end{table}
\end{center}

LP approaches provide significantly tighter bounds than second-order naive linear bounds and, surprisingly, allow for accurate extrapolations over more than an order of magnitude in sample size. However, these simulations assume a nearly infinite genome, and the convergence to this limit may be slow. Figure \ref{LvNn} shows the slow increase in prediction accuracy with sample size. In a sample with ten million polymorphisms, the 20-fold extrapolations are not very precise, but 8-fold extrapolations provide conservative lower bounds $4\%$ below the correct value and upper bounds $16\%$ above.

\subsection{Subsampling 1000 Genomes data}
\begin{figure}
\scalebox{0.77}{\includegraphics{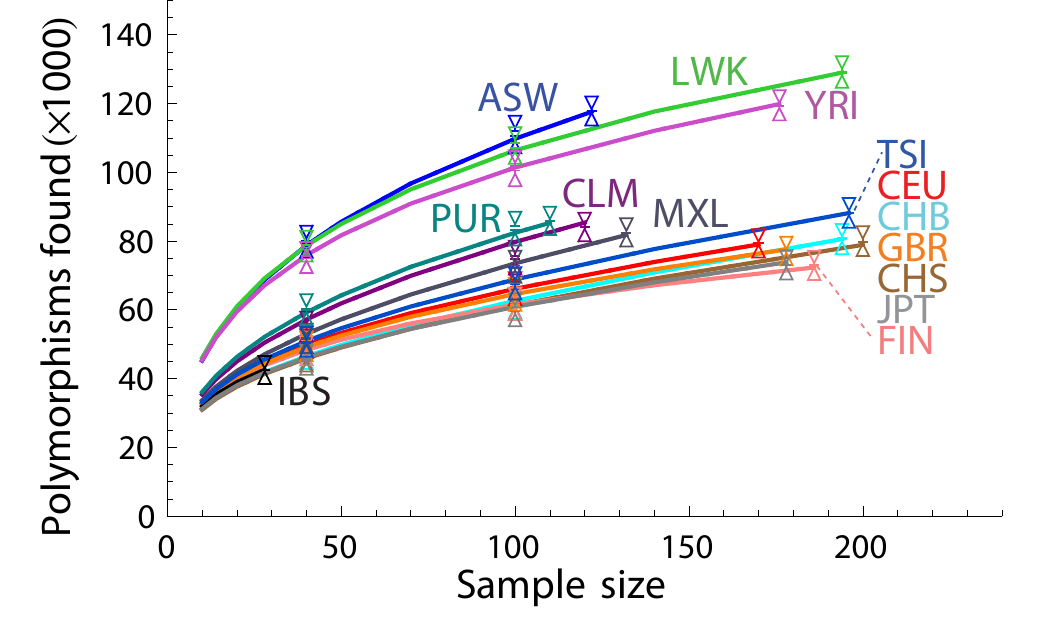}}
\caption{\label{subsample} Number of nonreference variants discovered for each of the 1000 Genomes Project  populations (solid lines). Linear programming (LP) predictions (shown as vertical intervals) are based on random subsamples of diploid individuals corresponding to 20, 40, and 100 haploid genomes. The triangle tips indicate the $95\%$ confidence maxima for the LP upper bounds and $95\%$ confidence minima for the LP lower bounds from 50 bootstrap runs. The short horizontal lines between triangles represent the width of the confidence interval for a single LP run; it is thinner than the line width in most instances. Each displayed interval uses the maximum subsample size available.  }
\end{figure}

The 1000 Genomes Project has released exome-capture data for 1092 individuals from 14 populations: some from predominantly European (CEU, TSI, GBR, 
FIN, IBS), African (YRI, LWK), East Asian (CHB, JPT, CHS) ancestry, and others of mixed continental ancestry (ASW, MXL, CLM, PUR). Figure \ref{subsample} shows the number of nonreference variants discovered as a function of sample size in each population.

\begin{figure}
\scalebox{.9}{\includegraphics{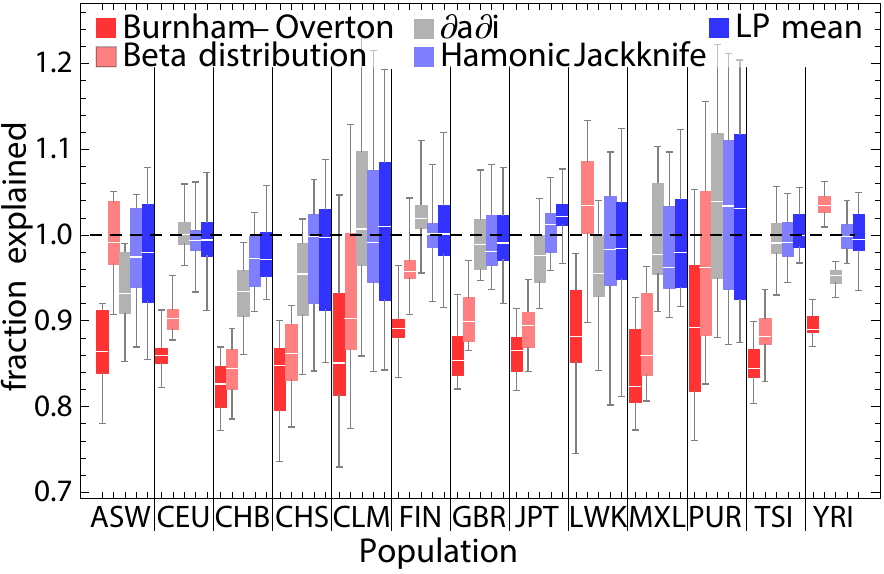}}
\caption{\label{subsampleboot}Predicted number of new variants discovered in $N=100$ haplotypes based on multiple subsamples of $20$ diploid individuals from 1000 Genomes populations, expressed as a proportion of the correct value. We display the existing  Burnham--Overton \protect\cite{Burnham:1979uv} and beta distribution \protect\cite{IonitaLaza:2009ik,IonitaLaza:2010jf} predictors; a prediction based on a 3-parameter demographic model fitted using $\partial$a$\partial$i; and the harmonic jackknife and linear programming (LP) approaches presented here.}
\end{figure}

To estimate the accuracy of the capture--recapture strategies, we randomly drew subsamples of 10, 20, and 50 diploid individuals, and extrapolated the number of discoveries from each subsample size to the next larger subsample size, or to the full population size. We find that the LP approach and the harmonic jackknife provide accurate estimates to within a few percent of the true values (Figures \ref{subsample} and  \ref{subsampleboot}), whereas the BO and beta distribution estimators underestimate the number of variants for most populations (Figure \ref{subsampleboot}). The demographic model approach is only slightly more biased than LP and hamonic jackknife, but it is also more intensive computationally and technically.
%he LP confidence intervals are wider than the jackknife confidence intervals because they also account for model uncertainty. In this case, model assumptions in the jackknife estimates help improve inference accuracy.

Even though the harmonic jackknife and LP approaches appear unbiased for all populations, the variance of the estimate depends on the population, with recently admixed populations (ASW, CLM, MXL, and PUR) showing the most variance, followed by populations with known cryptic relatedness (LWK and CHS). This variance indicates that the relatively small subsamples have ``personality" in these populations---if a sample contains an individual with particularly high European ancestry proportion, or a pair of closely related individual, it may sway the estimate in a way that would not occur in a more uniform sample. 
If we consider confidence intervals based on Poisson Random Field (PRF) parametric bootstrap, which assumes a perfectly homogeneous sample,  $95\%$ confidence intervals contain the observed data in $76\%$ of cases, whereas the harmonic jackknife confidence intervals contain the true value $68\%$ of the time  (see also Figure \ref{CItest}). If we exclude populations with admixture and relatedness, the proportion of confidence intervals containing the correct value increases to $92\%$ for LP and $86\%$ for the jackknife. Inhomogeneity effects are expected to decrease with sample size.

%  The $95\%$ PRF confidence bounds for linear programming contain the observed data in $76\%$ of cases, whereas the harmonic jackknife confidence intervals contain the true value $68\%$ of the time. This overconfidence of the PRF bootstrap confidence intervals is largely due to sample inhomogeneity. Populations that have received substantial recent admixture tend to have a high inter-individual variation in ancestry proportions (see, e.g., \cite{{Gravel:2012ip}}). This can be observed in Figure \ref{subsampleboot}, where the recently admixed populations ASW, CLM, MXL, and PUR exhibit substantially more variation across subsamplings. Among the remaining populations, LWK and CHS present by far the largest variation, consistent with the observation that these populations contain significant amounts of cryptic relatedness \cite{GenomesProjectConsortium:2012co}. Setting these populations aside, the proportion of confidence intervals containing the correct value increases to $92\%$ for LP and $86\%$ for the jackknife. PRF confidence intervals are therefore appropriate for homogeneous samples, and jackknife-based confidence intervals may be preferable for heterogeneous populations.

Importantly, both the harmonic jackknife and LP estimators appear to remain unbiased and accurate even for small inhomogeneous samples.  This is in stark contrast to the BO jackknife and the parametric beta distribution approach of \cite{IonitaLaza:2010jf,IonitaLaza:2009ik} \ref{subsampleboot}, which exhibit substantial bias for most populations.

 \subsection{Extrapolations using 1000 Genomes data} 
\begin{figure}
\scalebox{0.65}{\includegraphics{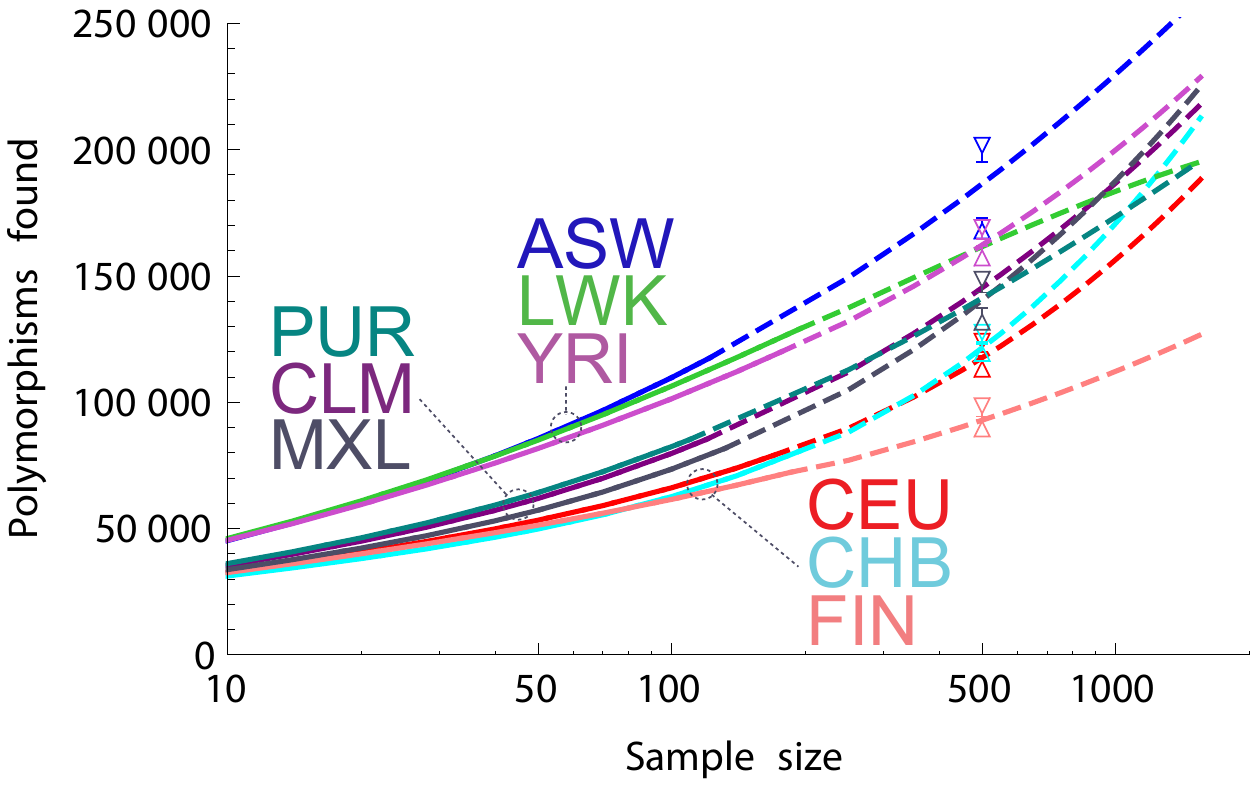}}
\caption{\label{1000pred} Predictions of nonreference exomic SNPs to be discovered in a selection of the 1000G populations as a function of the number of chromosomes sequenced, using the harmonic jackknife (dashed lines) and linear programming (LP; vertical intervals, shown for a subset of populations) on the full dataset. The triangle tips indicate the $95\%$ confidence maxima for the LP upper bounds and the $95\%$ confidence minima for the LP lower bounds from 50 bootstrap runs. The short horizontal lines between triangle tips represent the width of the confidence interval for the non-bootstrapped sample. }
\end{figure}

Extrapolations from the 1000G data are shown in Figure \ref{1000pred}. The harmonic jackknife and LP estimates are in good agreement. As in \cite{Nelson:2012cv}, we find that the African-American population (ASW), with predominantly West African and European continental ancestries,  has the highest predicted discovery rate. This is a joint effect of the high diversity of the African source population and of the contribution of two continental populations. By contrast, the Finns (FIN) show the least amount of diversity, consistent with a smaller recent effective population size.  Whereas the populations tend to cluster by continental ancestry at low sample sizes, reflecting shared histories, continental ancestry becomes less informative as sample sizes are increased, revealing consequences of the more recent histories of the sampled populations.   

 \subsection{The Exome Sequencing Project example}

\begin{figure}
 \scalebox{0.65}{\includegraphics{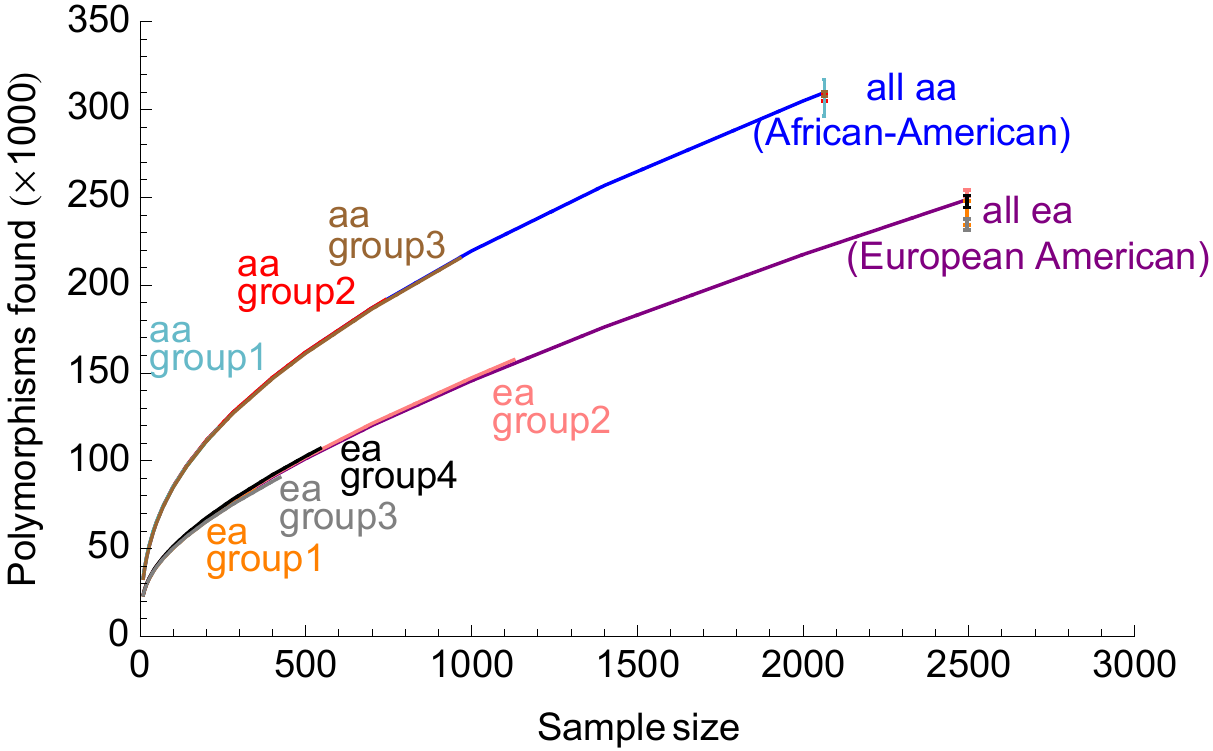}} 
\caption{\label{cohorts} Extrapolations based on different African-American (aa) and European-American (ea) sub-cohorts of the ESP meta-cohort on the full sample size. The different sub-cohorts correspond to data obtained from different projects and sequencing centres, as explained in the text. }
\end{figure}

To test whether the approach is applicable to cross-cohort prediction, we applied the method to data from the first 2500 sequenced individuals of the Exome Sequencing Project \cite{Tennessen:2012ck}, which combined data across different cohorts and sequencing centres. Figure \ref{cohorts} shows the total number of variants based on variants observed by four different sequencing groups (focusing on 1-LDL, 2-EOMI, 3-BMI \& EOS, and 4-lung diseases, see \cite{Tennessen:2012ck} for cohort and project descriptions). We find excellent agreement for predictions based on these subsets. The largest departure is from the European-American sample for group 3, which is also the smallest subset. 

\begin{figure}
 \scalebox{0.22}{ \includegraphics{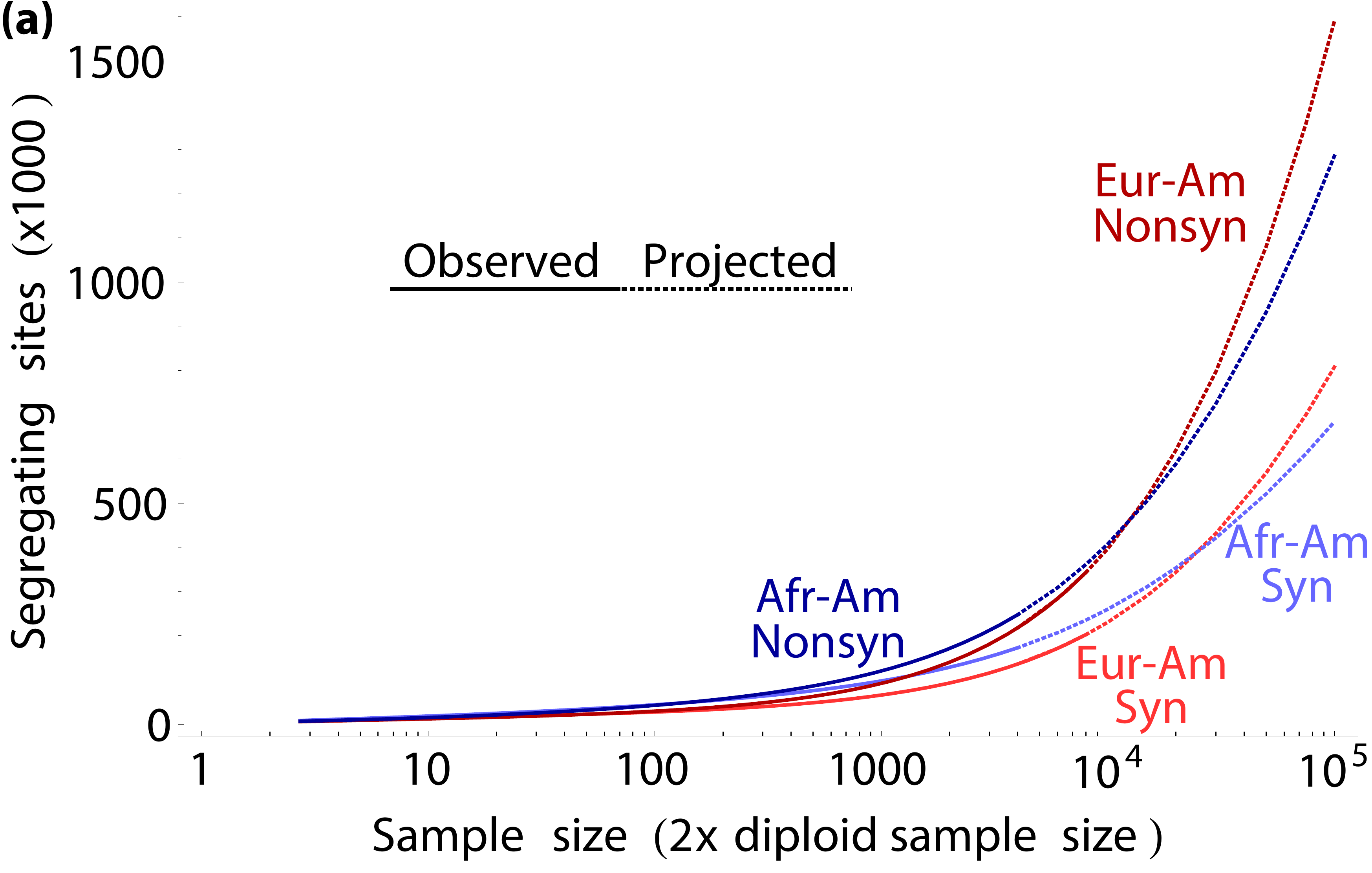}}

 \scalebox{0.3}{ \includegraphics{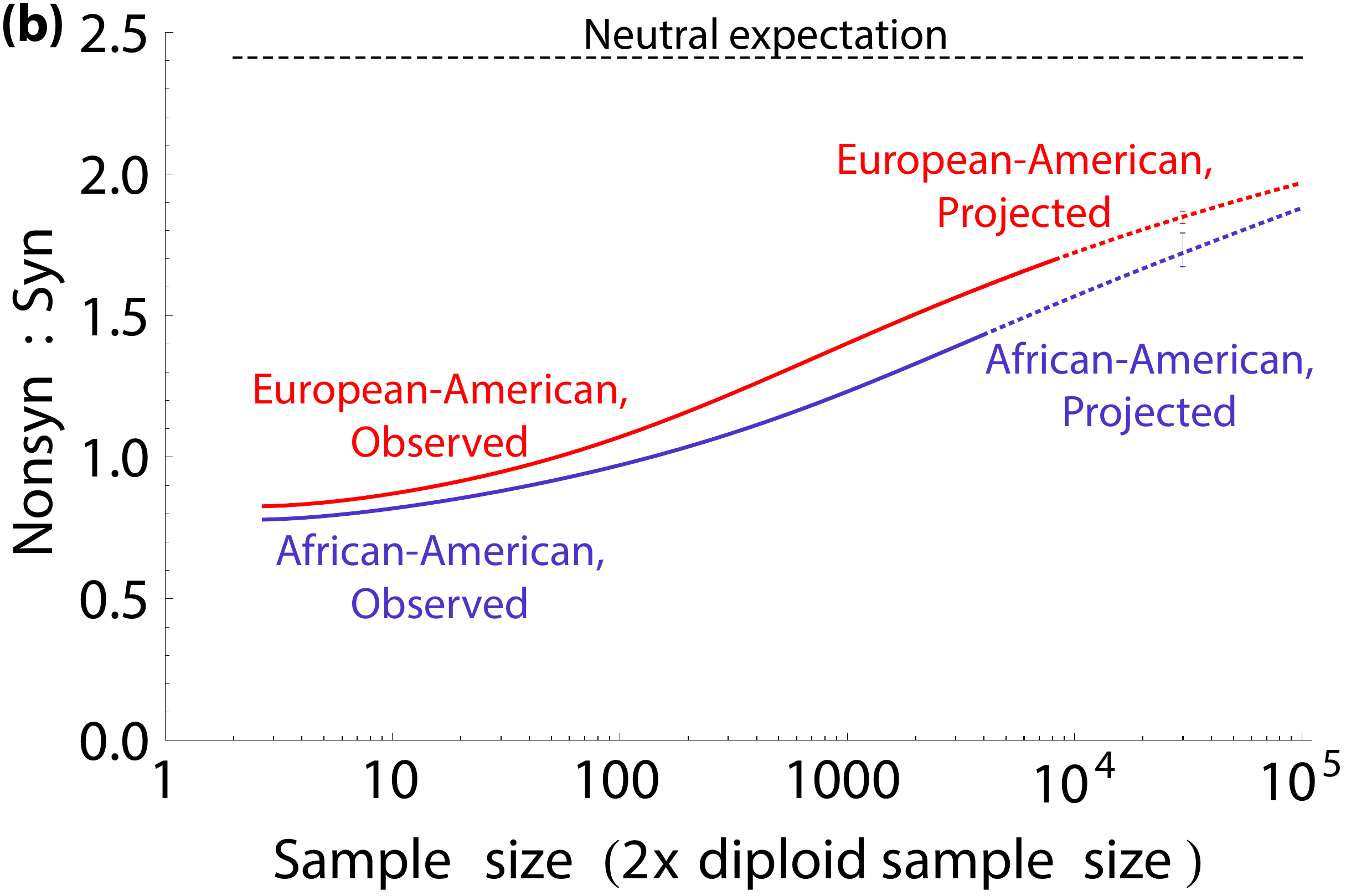}}

 \caption{\label{ESP} (A)  Projections for the number of synonymous and nonsynonymous sites in African-Americans and European-Americans based on the ESP sample (B) Observed and projected ratio of nonsynonymous to synonymous variants in the two populations.}
\end{figure}

Finally, to obtain the prediction for the largest possible sample, we considered the most recent data released by the ESP project, including over 6500 individuals of European-American and African-American descent, and generated predictions based on samples of 2000 African-Americans and 4000 Europeans, for sites with mean coverage above 40. Even though African-American populations have the most variable sites in present-day samples, we predict that this will no longer be the case in samples of $50,000$ diploid individuals, with $8.7\%$ of target sites predicted to be variable in European-Americans, compared to $7.2\%$ in African-Americans. The crossover is predicted to occur between 7,500 and 10,000 individuals. The predicted number of variants is higher in European-Americans for both synonymous and nonsynonymous variants (Figure \ref{ESP}(a)), but the proportion of nonsynonymous variants is likely to remain higher in Europeans than in African-Americans (Figure \ref{ESP}(b)), likely reflecting an excess of deleterious variants in Europeans.  The nonsynonymous:synonymous ratio will remain considerably lower than the neutral expectation under a Hwang--Green mutational model  \cite{Hwang:2004cf}  until samples in the millions are considered.

 \subsection{DNaseI footprinting}
\begin{figure}
 \scalebox{0.62}{ \includegraphics{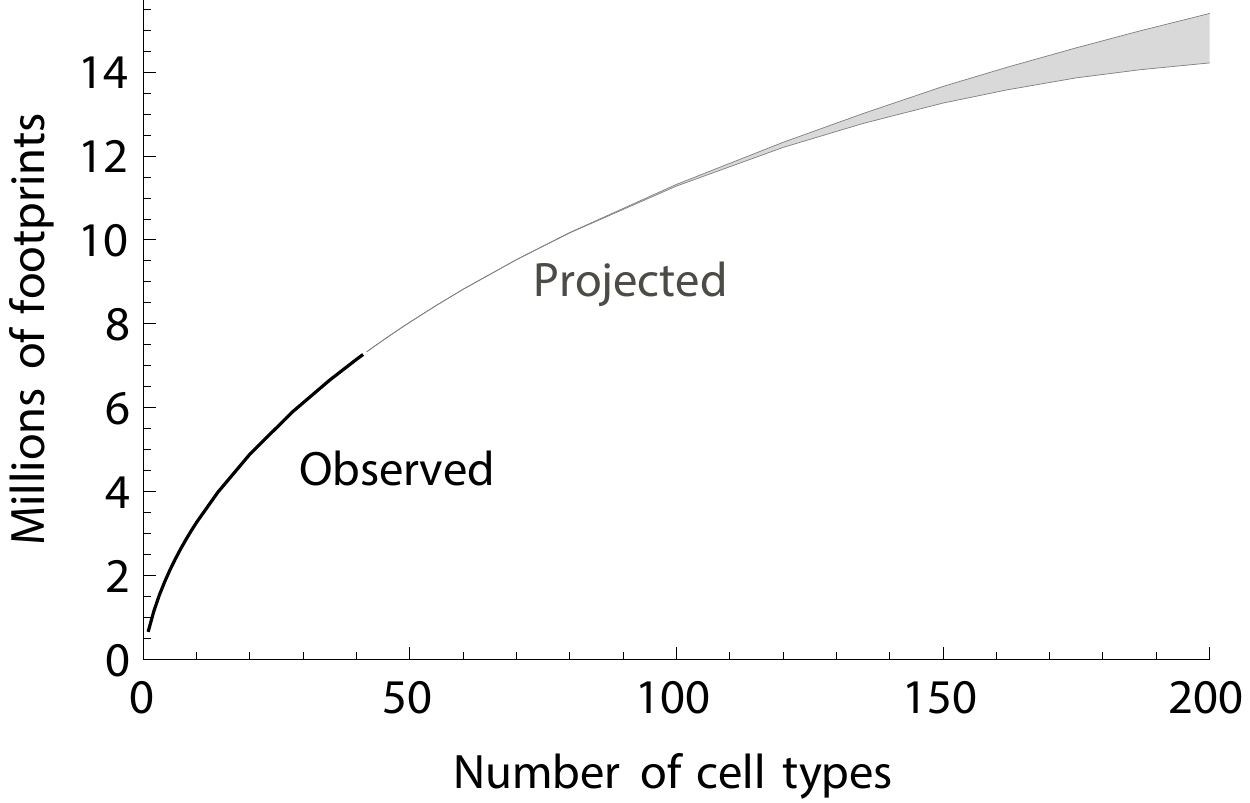}}
\caption{\label{DNAse} Observed and projected number of DNaseI footprints, marking putative transcription factor binding sites, as a function of the number of cell types studied. Projections use the LP approach, and the confidence interval represents the difference between the LP upper and lower bounds.}
\end{figure}

 Because the LP approach is nonparametric, it can be applied to any genomic feature that is present genome-wide and across samples. To illustrate this, we consider DNaseI footprints, which indicate sites where transcription factors bind to DNA and protect against cleavage by DNaseI. Encode produced a genome-wide map of such features across 41 different cell types \cite{Thurman:2012fe}. Using the same software, we are able to predict the number of transcription factor binding sites that will be identified as the number of cell types is increased. We identified sites as contiguous genomic regions where at least one cell type has a footprint. The LP bounds are particularly tight in this example (Figure \ref{DNAse}) , and the main source of uncertainty in this problem is the degree to which the choice of cell types in the Encode study is representative of the remaining cell types with respect to transcription factor binding. 
 
\section{Discussion}
\subsection{Theoretical  and statistical considerations}

Jackknife and LP approaches for finite and infinite extrapolation for the species-counting problem have been discussed before \cite{Efron:1976tt}. The sampling processes, binomial for t rabbit-counting problem, Poisson for the species-counting problem and hypergeometric in the genetics context, lead to fundamental differences. For example, in the Poisson case, an infinite number of data points is available because each species can be observed an arbitrary number of times. This allows for a (possibly divergent) formal expansion of the number of unobserved variants in terms of the $\{\phi(i)\}_{i=0,\ldots,\infty}$\cite{Efron:1976tt}. In the binomial and hypergeometric cases, we have only a finite number of observations $\{\phi(i)\}_{i=0,\ldots,n}$, making it clear that the series expansion cannot provide an exact result. In addition, the size $M$ of the population from which our sample was drawn determines how accurately we can perform extrapolations to sizes $N<M$, a situation that does not have a direct analog in the Poisson case. 

A difference between the genetics problem and both the species- and rabbit-counting problems is the target extrapolation size: in many ecological problems, the number of field trips itself is not a variable of interest, and the ultimate goal is to extrapolate to infinite sample sizes. In such a case, the resulting confidence interval would be semi-infinite.  Intuitively, we can never exclude the possibility that a very large number of very uncatchable rabbits have eluded detection. As a result, all point estimates require implicit or explicit assumptions about the existence of such sneaky rabbits. This led to the correct statement \cite{Link:2003wo} that nonparametric point estimates are impossible in the rabbit-counting problem. Nonparametric point estimates are still impossible in the finite extrapolation context studied here: there is a finite interval of values equally consistent with the data, and any choice implies parametric assumptions.   However, if this finite interval is narrow enough, we may not need point estimates: in many cases, the predicted consistency interval is narrower than other uncertainty sources. Thus nonparametric point estimates do not exist, but this may not be important: LP provides a practical, nonparametric interval estimator.   

Some of the strategies that we have proposed may translate back to the ecology problems. One example is the coarsening strategy used in the finite genome problem, in which we merge bins of less-informative common variants to improve computational performance and accuracy. We have found that extrapolations can be accurate beyond 20-fold increases in sample size, a finding surprising in the light of previous work. The accuracy of projections as a function of sampling scheme, sample size, and model assumptions remains a largely open question of considerable theoretical interest.

We have discussed five different extrapolation strategies in this article, and found that two of these (the harmonic jackknife and LP) outperformed the others (beta distribution, demographic modeling, and BO jackknife). The beta distribution and demographic modeling suffer from their attempt to model the entire allele frequency distribution via a few-parameter family of models. With larger datasets, departures from these model families become more significant and lead to the observed biases. By contrast, the jackknife approaches fit a similar number of parameters but model only the rare end of the frequency spectrum, which contains most of the information about future discovery rates. In that sense, they make better use of fitting parameters, but the assumptions of the BO jackknife differ too much from realistic genetic scenarios. The assumptions of the harmonic jackknife, by contrast, include realistic genetic scenarios, and as a result the extrapolations are quite accurate. Finally, linear programming does not require any assumptions about allele frequency distribution, and as a result is much more broadly applicable than the other methods. Furthermore, in the infinite-genome limit, it uses all the information available in the data, and we have found it to be surprisingly accurate. Thus, the nonparametric and less parametric methods fare very well in this comparison. This is because the large dataset is very informative about the underlying distribution, making parametric assumptions both less useful and more risky.

\subsection{Practical aspects}

Among the many approaches that we have discussed for predicting the number of unobserved variants, linear programming and the harmonic jackknife stand out as being less biased. So which one should be used in practice? The harmonic jackknife approach is the fastest to compute, and provides comparable results to LP for a diversity of realistic evolutionary scenarios. When applicable, the parametric assumption may help obtain slightly more precise results compared to the linear programming approach. However, we cannot exclude the possibility that it will perform poorly for strong departures from the Standard Neutral Model,  or in systems with entirely different dynamics, such as transcription factor binding sites across cell types.  By contrast, the LP approach does not assume a specific form for the distribution of allele frequency, and it can easily be modified to entirely different problems, or account for additional (linear) constraints with little to no validation effort required. For example, we could easily have imposed a constraint that the underlying frequency spectrum be strictly decreasing over some range of frequencies, leading to a narrower confidence interval. Implementing such a change in the jackknife approach would have been challenging. In many cases, LP is the only trustworthy solution. In a few cases of very large samples, LP problems may require additional optimization and it may be easier to use the jackknife. In cases where both are applicable, we suggest using both methods; if the jackknife falls outside the LP bounds, we know that its assumptions were not met, and the LP estimator should be used. Otherwise, the jackknife estimator is probably the most principled guess among the values allowed by LP. 

Where computationally tractable, the linear programming approach has important advantages, the main one being the easy transferability to different types of problems. However, from a practical standpoint, jackknife estimators are not to be discounted. They are extremely fast and, even though the underlying assumptions may be difficult to interpret in terms of the fundamental processes involved, they tend to produce accurate estimators in a wide range of scenarios. Comparison of the exact and jackknife weights (Figure \ref{jkapprox} and the Appendix) provides good intuition for this relative robustness. Finally, even though the LP bounds are asymptotically optimal among nonparametric estimators, a visual inspection of the underlying distributions (Figure \ref{minmax}) suggests that even fairly conservative biological assumptions can produce narrower bounds. For example, requiring that the large population be drawn from an even larger population resulted in improved intervals (Table \ref{tabex}). Some other assumptions, such as smoothness or monotonicity over a range of frequencies, can easily be accommodated in a linear program and would be worth exploring. 

The most crucial assumption underlying the extrapolation methods presented here is random sampling---we must be able to consider the existing sample as a random subset of the larger population. By contrast, we found that recent admixture, population structure, linkage, and cryptic relatedness do not seem to cause substantial biases, and the LP approach should be applicable to datasets whose evolution is fundamentally different from that of SNPs. We found that some of these factors change the variance of these estimates: populations with more sample inhomogeneity and cryptic relatedness lead to more variable estimates, but we expect these effects to decrease when the sample size is increased. We do not expect linkage disequilibrium (LD) to bias our estimates, because LD does not affect the expected frequency spectrum that is the starting point of our estimates. Furthermore, we are mostly concerned with rare variants, which typically are not in high LD with each other. Thus, both the expectation and variance of genome-wide estimates should be little affected by LD. There may be applications where variances are more affected by correlations: in the transcription factor binding example, we may imagine that cell-type-specific transcription factor binding sites cluster, in which case the Poisson Random Field that we used to estimate confidence intervals may become a poor approximation. In such cases, leave-one-out experiments should be performed to assess confidence intervals.

The random subsampling assumption remains a demanding one---in practice, subtle differences in sampling make it likely that results extrapolated from one sample will not apply to another one.  Witness the 1000 Genomes data (Figure \ref{1000pred}), which sampled largely distinct populations. In this case, very different discovery rate estimates reflect the different recent histories of the populations. On the other hand, we also find that results in the large medical cohorts from ESP are exquisitely reproducible across cohorts, even though these are definitely not subsamples of each other. By contrast with the 1000 Genomes data, the ESP meta-cohort was assembled using comparable (even sometimes overlapping) cohorts \cite{Tennessen:2012ck}. This emphasizes how the methods presented here are applicable to make predictions across panels that are similar but not identical.  

%We finally point out that even though the success of the nonparametric projection methods presented here is encouraging for experiment design, it should also serve as a caveat for population geneticists wishing to test demographic and selective models. If discovery rates are largely determined by sampling statistics, they are not very informative about specific models.   

%>>> foo=open('DesignatedAncestry.txt','r')
%>>> same=0
%>>> diff=0
%>>> lines=foo.readlines()
%>>> lsps=map(lambda i: i.split(),lines)
%>>> for lsp in lsps[1:]:
%...     if lsp[1]==lsp[2] and lsp[1] is in ['WhiteorEuropeanAmerican','BlackorAfricanAmerican']:
%  File ``<stdin>", line 2
 %   if lsp[1]==lsp[2] and lsp[1] is in ['WhiteorEuropeanAmerican','BlackorAfricanAmerican']:
  %                                   ^
%SyntaxError: invalid syntax
%>>> for lsp in lsps[1:]:
%...     if lsp[1]==lsp[2] and lsp[1] in ['WhiteorEuropeanAmerican','BlackorAfricanAmerican']:
%...             same+=1
%...     elif lsp[1]!=lsp[2] and lsp[1] in ['WhiteorEuropeanAmerican','BlackorAfricanAmerican'] and lsp%[2] in ['WhiteorEuropeanAmerican','BlackorAfricanAmerican']:
%...             diff+=1
%... 
%>>> same
%2263
%>>> diff
%16

Large sequencing efforts such as the 1000 Genomes project often start with a pilot project aimed at testing the technology, identifying possible issues, and providing funding bodies and stakeholders a preview of the full project. The methods presented here provide a straightforward and well-calibrated approach to estimating a key deliverable in the final project. As the project is completed, the results can be compared to the initial predictions, assessing the impact of methodological and sampling changes between the pilot and the main phase. Of course, the final results can be extrapolated to serve as a baseline prediction for the next set of experiments. 

Predicting the number of variants to be discovered in a new sample is one of the few areas where population geneticists studying long-lived organisms can make experimental predictions, and as such is an important tool for population genetics hypothesis validation. The success of the nonparametric methods presented here shows that this can be performed to high accuracy. However, the success of nonparametric methods and their robustness to linkage, demography, population structure, and selection suggests that accurate model-based predictions of future discovery rates do not necessarily provide additional evidence that these effects are correctly taken into account. Over-fitted models that are consistent with the data should provide predictions within the LP confidence intervals. Model-based predictions should therefore improve upon the LP predictions to validate the model. By contrast, the LP prediction provides a strong test of whether the initial sample can be considered a random subsample of the full population, a commonly used assumption in population genetics models. This work therefore demonstrates that nontrivial falsifiable predictions can easily be generated and tested against future genomics experiments. I hope that it will encourage more genomicists to put their head on the prediction block.

\section{Acknowledgements}
The authors wish to acknowledge S. Baharian, R. Gutenkunst, J. Kelley, O. Cornejo, S. Shringarpure, and M. Carpenter for useful comments; E. E. Kenny and C.D. Bustamante for discussions and help with data access; the support of the National Heart, Lung, and Blood Institute (NHLBI) and the contributions of the research institutions, study investigators, field staff and study participants 
in creating this resource for biomedical research.  Funding for GO ESP was provided by NHLBI grants RC2 
HL-103010 (HeartGO), RC2 HL-102923 (LungGO) and RC2 HL-102924 (WHISP). The exome sequencing 
was performed through NHLBI grants RC2 HL-102925 (BroadGO) and RC2 HL-102926 (SeattleGO).

%\processdelayedfloats 

%\bibliographystyle{plain}
%\bibliographystyle{mychicago}
%\bibliography{cap_recap}

\pagebreak
\setcounter{page}{1}
%\section*{Appendix}

\subsection*{Jackknifes and naive linear bounds}
\label{finite}
We can obtain both upper and lower bounds for the number of undiscovered variants by linear combinations of the $\phi(d)$.  
%$$V(N)-V(n)=\int_0^1\left((1-f)^n-(1-f)^N\right)\Phi(f) df$$
To do this, we note that the equations for the number of missed variants
$$V(N)-V(n)=\int_0^1\left((1-f)^n-(1-f)^N\right)\Phi(f)df$$
and for the number of variants at a given allele frequency

$$\phi_n(j)=\int_0^1 {n \choose j} f^j (1-f)^{n-j} \Phi(f) df$$
have a very similar form. The only difference is a `weight factor' before $\Phi$. If the weight function $w_{n,N}(f)=(1-f)^n-(1-f)^N $ can be approximated by functions of the form $b(f,\vec{\alpha})=\sum_{i=1}^d \alpha_i  f^i (1-f)^{d-i}$ then we can approximate  $V(N)-V(n)$ in terms of the observed $\phi_n(i)$. In fact, this is exactly what the jackknife estimates do--A jackknife estimator would correspond to a function $$J(f) = \sum_{i=1}^d  \beta_i \phi(i),$$ with the $\vec \beta$ chosen such that 
$V(N)-V(n)=\int_{0^+} w_{n,N} (f)\tilde \Phi(f),$ for a particular $d$-parameter family of models $\tilde  \Phi(f)$, thought \emph{a priori} to be a reasonable proxy for $\Phi(f)$. This interpretation of the jackknife provides intuition about the behavior of jackknife estimators when the underlying model is not within $\tilde \Phi(f)$; comparison of the jackknife weight $J(f)$ and the correct weight $w(f)$  (Figure \ref{jkapprox}) provides an idea of the general accuracy of the jackknife estimate, and an idea of the frequencies that are more (or less) sensitive to errors.

However, we can also use the similarity between the expressions to obtain strict bounds on $V(N)-V(n)$, by choosing functions  $b(f,\vec{\alpha})=\sum_{i=1}^d \alpha_i  f^i (1-f)^{n-i}$ that are strict bounds to $w_{n,N}(f)$. The best such bounds will be attained when the approximating function $b(f,\vec{\alpha})$ touches but does not cross $ w_{n,N}(f)$

 We can show that the best upper bound with $d=2$ is $V(N)-V(n)<(N/n-1) \phi(1).$ There is a one-dimensional family of lower bounds which are optimal for at least one function $\Phi(f),$ parameterized by the contact point $0\leq f_0\leq 1$ where    
\begin{equation} 
\begin{split}
b_2(f_0, \vec\alpha_{f_0})=w_{n,N}(f_0)\\
b'_2(f_0,\vec\alpha_{f_0})=w'_{n,N}(f_0).
\end{split}
\end{equation}

To see that these $\vec\alpha_{f_0}$ exist and define lower bounds, consider the first, second, and third derivatives of the function $\frac{w_{n,N}(f)-b(f,\vec{\alpha})}{(1-f)^{n-2}}.$
  
For each $f_0$, we can solve for $\vec \alpha_{f_0}$, and thus obtain a lower bound to 
$V(N)-V(n)$. Given a sample, one can calculate all bounds and use the tightest. Figure \ref{flower} and Table \ref{tabex} show results using this approach with simulated data. It is easy to derive bounds with higher $d$, but the process of establishing the optimal bound is more challenging. Extrapolations based on upper bounds with  $d=3$ are shown on Table \ref{tabex}.  

 As in the case of jackknife estimates, higher order for the bounds means reduced bias, but also reduced stability in the presence of errors.

\subsection*{Known proportion of invariant sites}

In the ecology problem, the proportion of individuals or species that have not been observed is unknown; it is the object of the inference. In the genetic context, the total number of sequenced sites $L$ may be known; the object of the inference is to determine the proportion of these sites that would be variable in a larger sample. This does not fundamentally change the inference process: %Furthermore, it simply modifies the LP estimator $\hat V(N)$ into $\min(L,V(N))$.
\subsubsection*{Jackknife bounds}

In the jackknife case, we are provided with one additional function $(1-f)^N$ to try to obtain a linear bound to the weight functions $w_{n,N}(f)$. In the infinite-extrapolation case ($N=\infty$), we now have an upper bound to the number $U$ of undiscovered variants: $U\leq \phi(0)$. This is an inequality because variants with frequency $0$ are counted in $\phi(0)$ but not in $U=\int_{0^+}^1(1-f)^n\Phi(f)$. 

Finite extrapolation bounds can be improved using the knowledge of $\phi(0)$, by following the procedure described in the `Naive linear bound' section for the optimization of the $\vec \alpha_i.$ However, we do not study these in detail here.

\subsubsection*{Linear programming bounds} 
\label{sJKbounds}

In the linear programming framework, the observed $\phi(0)$ is easily incorporated as an additional equality constraint stipulating that $\sum_i \Phi(i)=\sum_j \phi(j)$. Intuitively, we expect that the additional constraint will help narrow the confidence interval. %Interestingly, when the total sample size is equal to the extrapolation size (i.e., $M=N$), knowledge of the fraction of variable sites does not allow for non-trivial improved bounds; the best lower bound remains the same, and the best upper bound is the minimum between the upper bound as described above and the total number of sites. This is 

However, when the total sample size is equal to the extrapolation size (i.e., $M=N$),  this provides limited information because the additional constraint involves a new variable, $\Phi(0)$, that is not involved in the objective function $V(N)$. Thus, $\Phi(0)$ can be adjusted to satisfy the constraint without affecting $V(N)$. Starting from a vector $\Phi^*(i)$ realizing the upper bound $V^{*}_\uparrow(N)$ for the problem with $\phi(0)$ unknown, such an adjustment is possible unless $\sum_{i=1}^N \Phi^*(i)>\sum_{d=0}^n \phi(d),$ in which case $\Phi(0)$ would be negative, violating the constraint  $\Phi(0)\geq 0.$ In such a case, convexity ensures that the optimal solution must satisfy $\Phi(0)=0$, and $V_\uparrow(N)=\sum_{d=0}^n \phi(d)$. Thus, in general, we simply have the somewhat disappointing result  $V_\uparrow(N)=\min\left(V^{*}_\uparrow(N),\sum_{d=0}^n \phi(d)\right).$ The same argument holds for the lower bound, but since $V^{*}_\downarrow(N)\leq \sum_{d=0}^n \phi(d)$, the lower bound is unchanged by the additional information.

This argument does not hold if the population size $M$ is larger than the extrapolation size $N$ because, in that case,  $\Phi_M(0)=0$ does not imply $V(N)= \sum_{d=0}^n \phi(d)$. Indeed, we find an improvement of the upper bound that becomes more pronounced as the number of invariant site in the sample of size $M$ is decreased.

\subsubsection*{Jackknife equivalence} 
\label{jkequiv}

We wish to show that the jackknife expansions A:  $V(N)-V(n)=\sum_{i=1}^p a_i \left(H(N)-H(n)\right)^i$, and B: $V(N)-V(n)=\sum_{i=1}^p b_i H^i(N)-H^i(n)$ lead to the same predictions. Both expansions can be written in the third expansion form C:  $V(N)-V(n)=\sum_{i=0}^p c_i(N) H(n)^i$, for different parameterizations of $c_i(N)$. Importantly, these parameterizations do not involve $n$. In the parameter estimation, we use in the three cases the constraints $V(n)-V(n-j)=\sum_{i=0}^p  c_i \left(H^i(n-1)- H^i(n)\right),$ for $j=\{1...p\}.$ These provide $p$ equations for $p$ unknowns $\{c_i\}_{i\geq1}.$ We can solve for these independently of $N$. We could equally well expand the $c_i$ in terms of, say,  the $a_i$, solve a linear equation for the $a_i$, and substitute these back to produce exactly the same expansion. Thus, the expansions A, B, and C are equivalent for $i>0.$   In expansion C, the dependence on $N$ enters only after we impose that $V(N)-V(n)$ must be zero when $N=n$. This imposes $c_0=-\sum_{i=1}^p i c_i H(N)^i$. This simple form of the estimator, made explicit in expansion $B$, was obscured by the poor parameterization choice of expansion $A$: whereas the $\{b_i\}_{i\geq 1}$ depend only on $n$, the $\{a_i\}_{i\geq 1}$ are messy functions of $N$ and $n$.

%\end{article}

%\bibliographystyle{genetics}
%\bibliography{cap_recap}
%\bibliography{allpapers2}

 % \pagebreak

\subsection*{Supplementary figures}
\numberwithin{figure}{section}

\makeatletter \renewcommand{\thefigure}{S\@arabic\c@figure} \renewcommand{\thetable}{S\@arabic\c@table} \makeatother 

\setcounter{figure}{0} 
\setcounter{table}{0}

%\begin{figure}[h!]
%\scalebox{0.6}{\includegraphics{./figures/BOJK}}
%\caption{\label{BOJK} Distribution of Z-scores of the known variants count among Poisson Random Field bootstrap distribution of estimates based on subsamples.   }
%\end{figure}

%\begin{figure}[h!]

%\caption{\label{BO40} Distribution of prediction accuracy based on different subsampling of 1000 Genomes data, under the assumptions of the Burnham-Overton jackknife \cite{Burnham:1979uv} . We show the distribution of $95\%$ confidence low (L) and high (H) bounds for variants to be discovered in 50 diploid individuals based on subsamples of 20.  All values are normalized by the correct value. For each subsample, upper and lower bounds are obtained under a Random Poisson Field approximation. The BO assumption underestimates the total number of variants by a few percents }
%\end{figure}

\begin{figure}[h!]
\scalebox{0.51}{\includegraphics{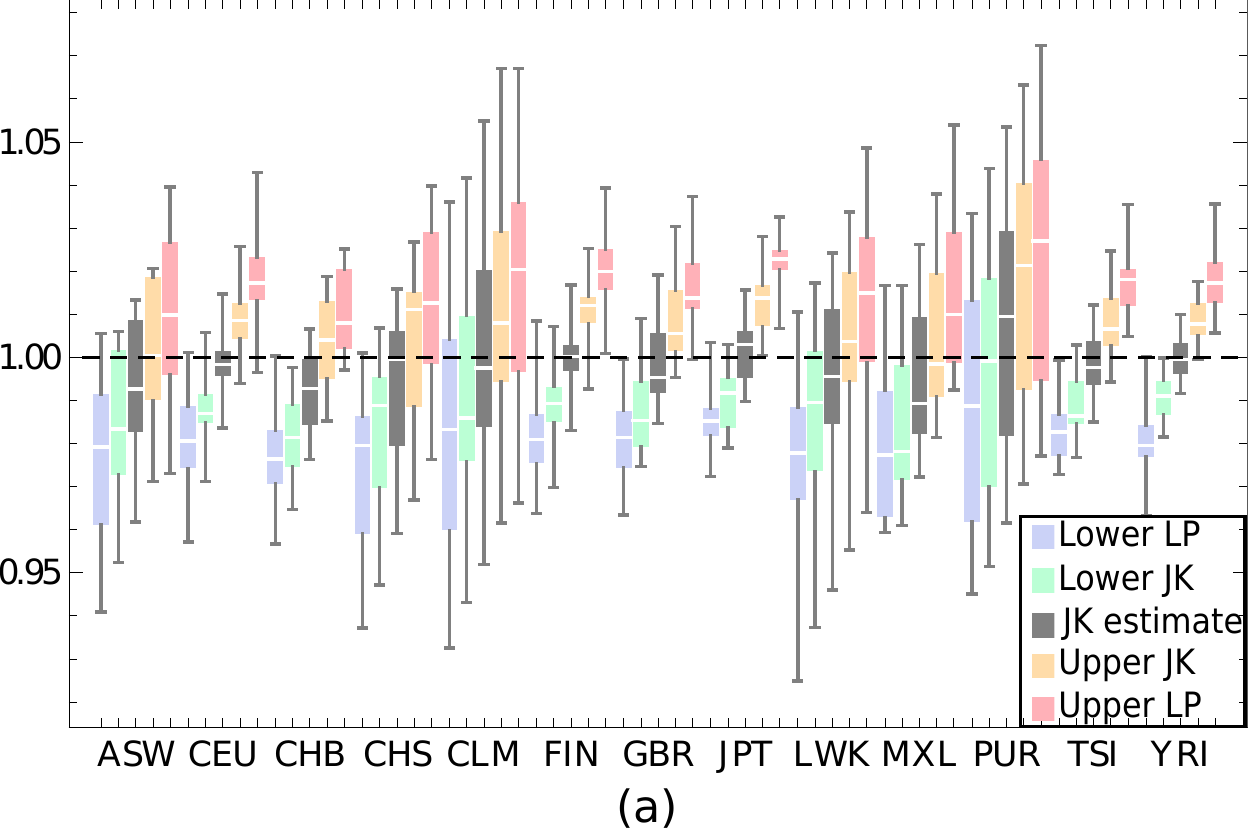}}
%\scalebox{0.6}{\includegraphics{./figures/JKBO40_200}}
%\scalebox{0.8}{\includegraphics{./figures/BETA40_200_v2_illust.pdf}}
\caption{\label{CItest} Distribution of predictions for $N=100$ based on multiple subsamples of 20 diploid individuals from 1000 Genomes populations, expressed as a proportion of the correct value. We display the jackknife prediction, and upper and lower $95\%$ bootstrap confidence intervals based on the Jackknife estimator and Linear Programming. Recently admixed populations (ASW,CLM,MXL,PUR), and populations with cryptic relatedness (ASW,CHS,MXL,LWK) show more variation across sub-samples, reflecting sample heterogeneity.}
\end{figure}

\begin{figure}[h!]
\scalebox{0.6}{\includegraphics{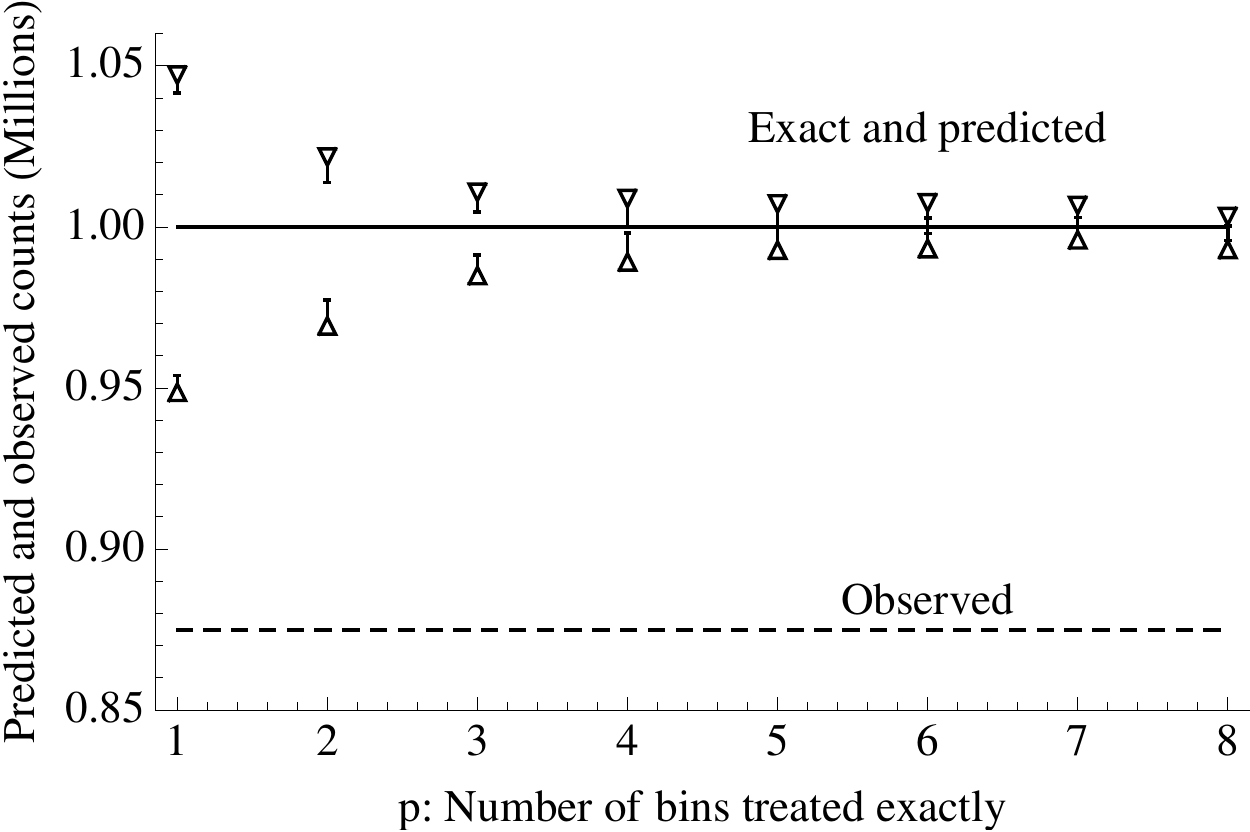}}
\caption{\label{lpbybin} Linear Programming upper and lower bounds, extrapolating from 50 chromosomes sampled from a population of 100 chromosomes containing 1 Million SNPs following the frequency distribution from Equation \eqref{simu}. The sample was generated assuming Poisson noise in each bin. Upper and lower bounds are calculated for 20 different Poisson resamplings of the sample, and $95\%$ confidence intervals were obtained (vertical lines). The tips of the upwards and downwards pointing triangles represent the $95\%$ confidence intervals of the lower and higher bounds, respectively.  LPs with $p\geq 9$ were not feasible.  The `observed' line represents variants observed in the sub-sample. }
\end{figure}

\begin{figure}[h!]
\scalebox{0.6}{\includegraphics{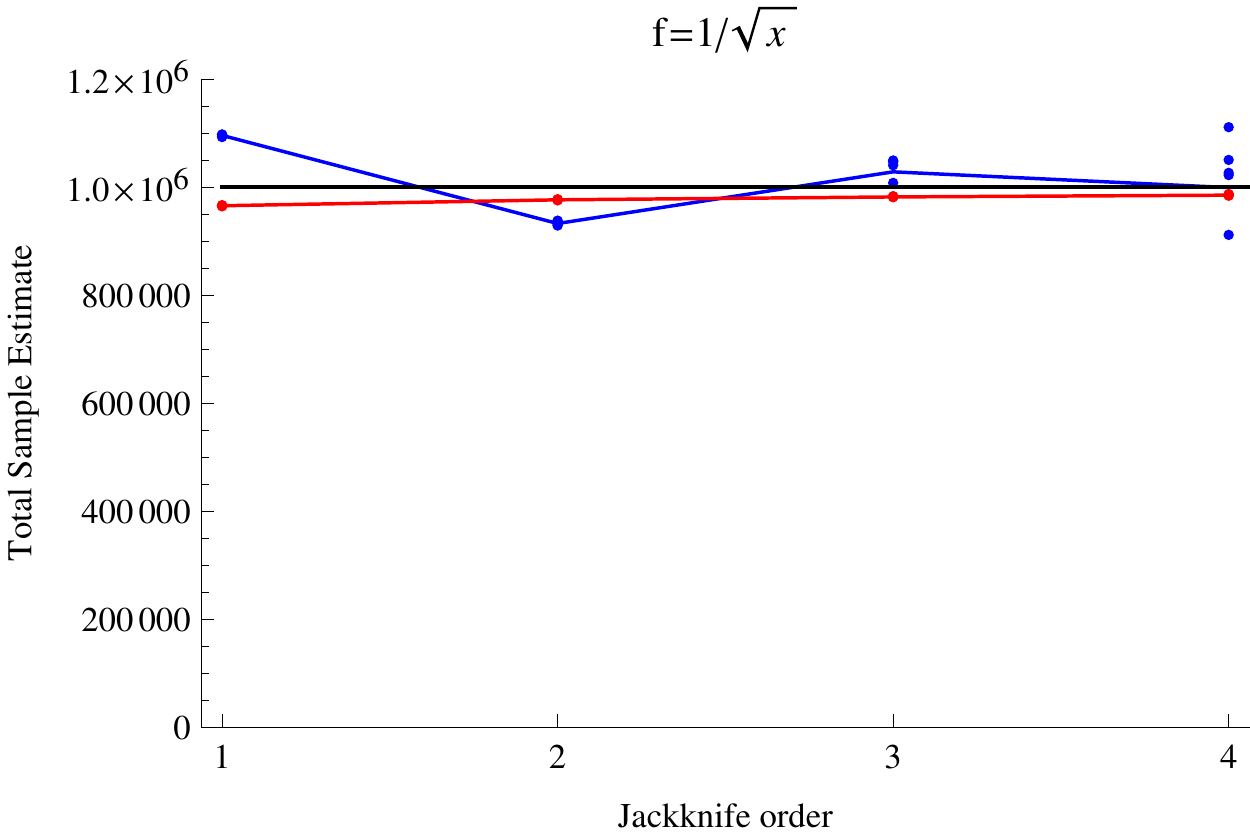}}
\scalebox{0.6}{\includegraphics{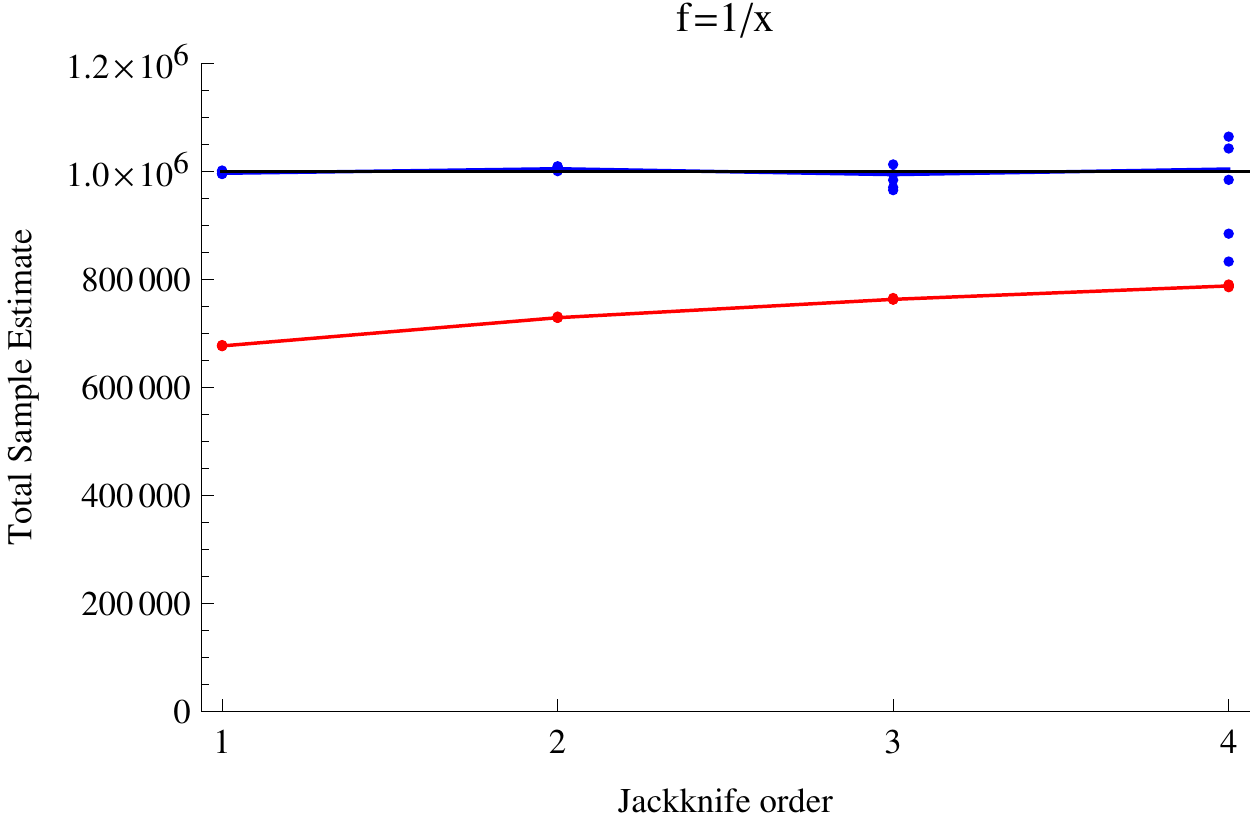}}
\scalebox{0.6}{\includegraphics{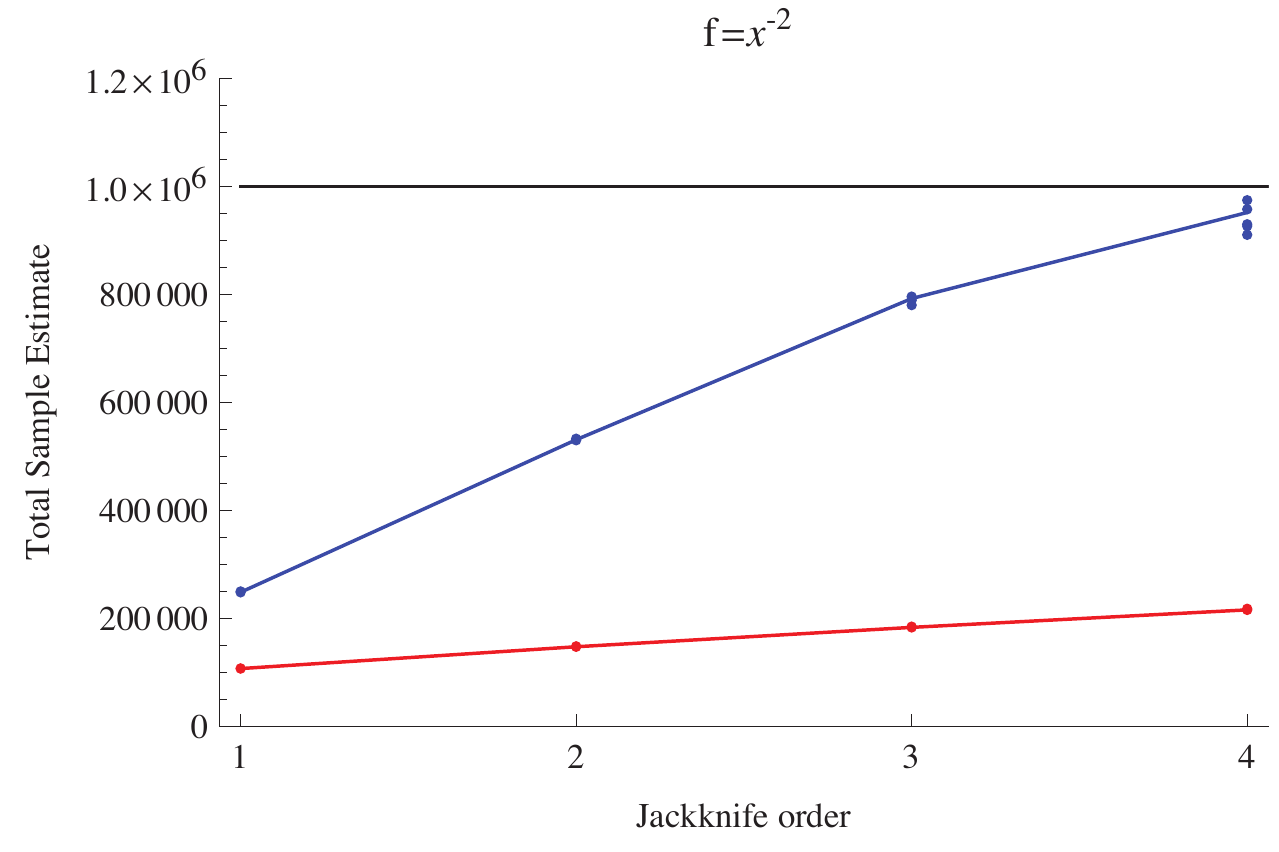}}
\caption{\label{SQRT} Jackknife simulations using the BO assumptions (Red) and the harmonic assumptions (Blue) for three different functional forms of the site-frequency spectrum, extrapolating from a population of 100 to 5000 chromosomes, based on a total count of 1,000,000 SNPs. The middle panel corresponds to the Standard Neutral Model. }
\end{figure}

\begin{figure}[h!]
\scalebox{0.6}{\includegraphics{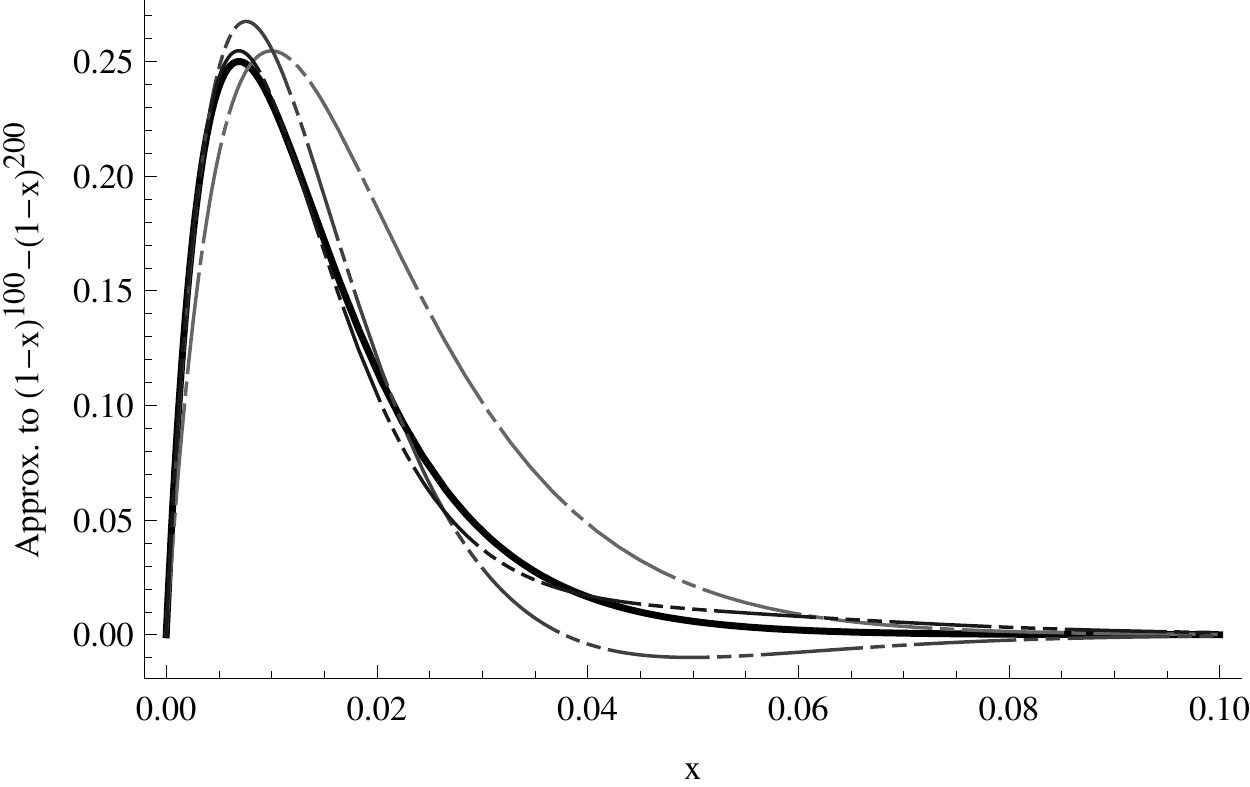}}
\scalebox{0.6}{\includegraphics{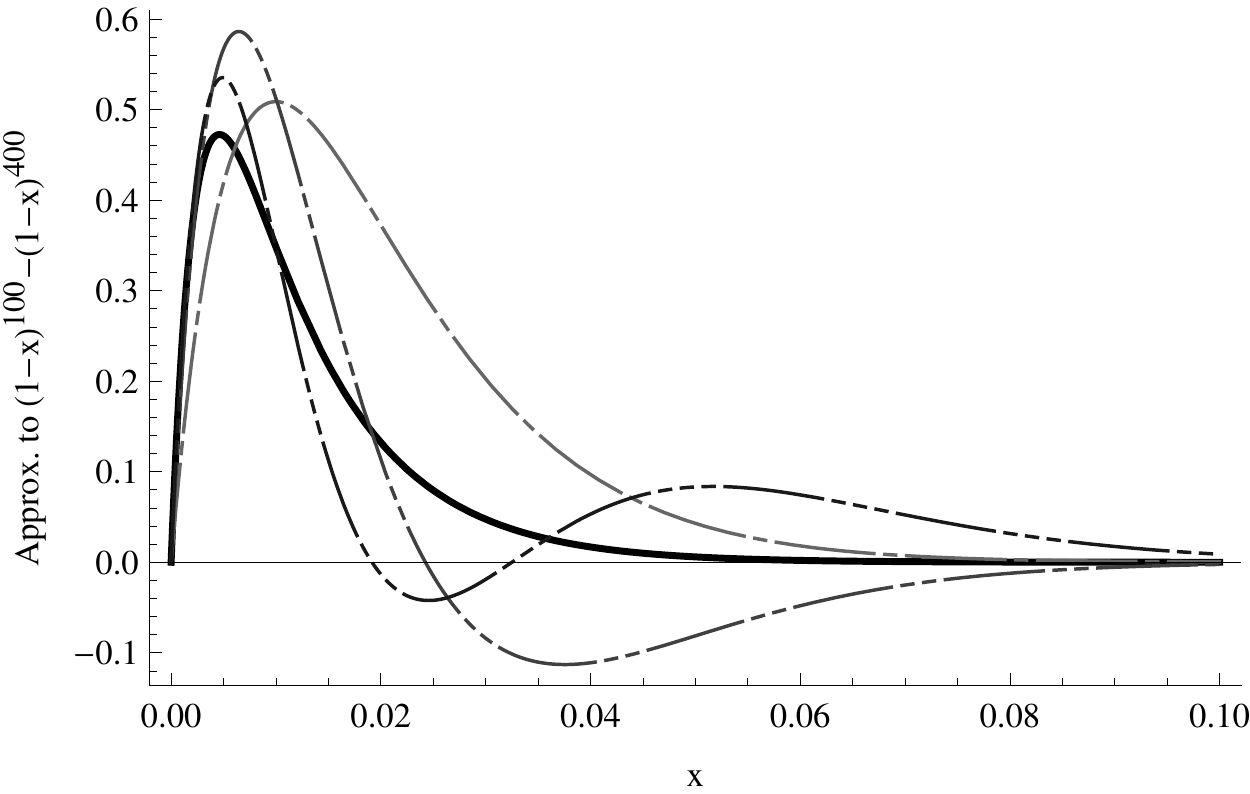}}
\scalebox{0.6}{\includegraphics{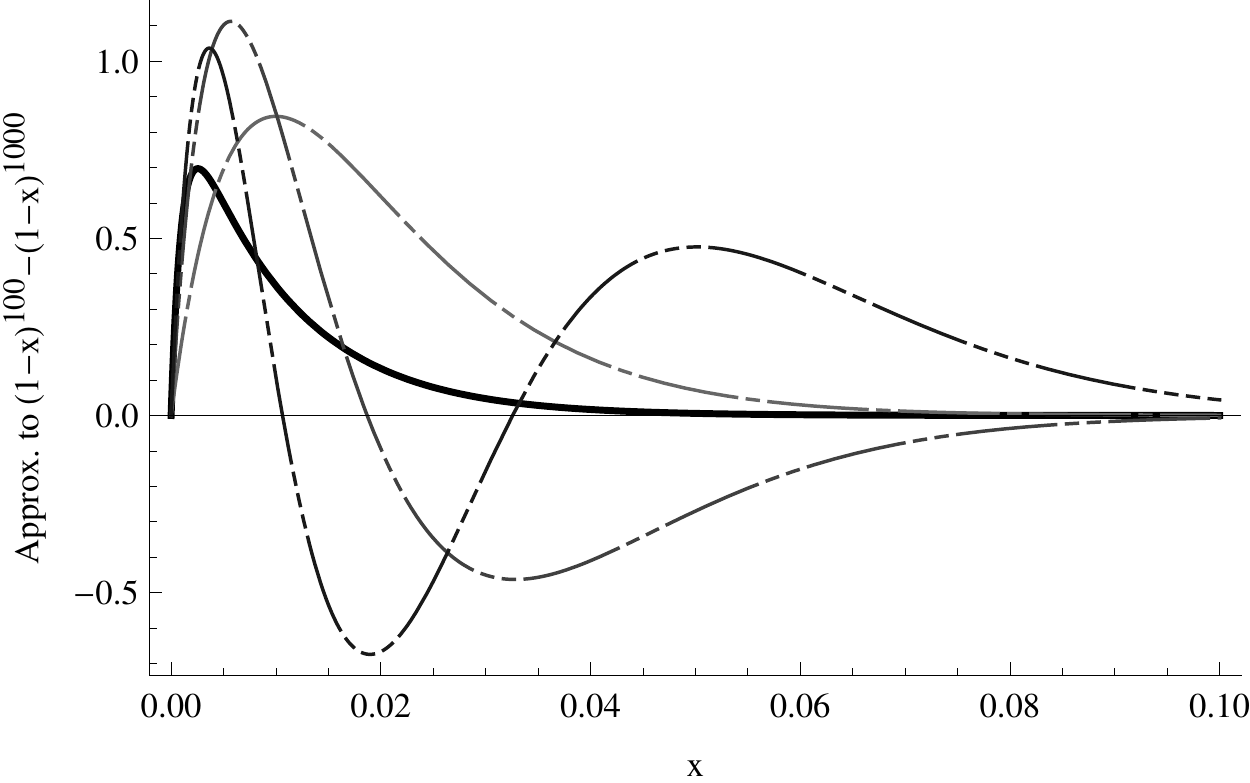}}
\caption{\label{jkapprox} Comparison of the true weight $w(x)=(1-x)^n-(1-x)^N$ used in the infinite-genome expression \eqref{basic} for the number of missed variants (thick solid line) to the jackknife approximate weights (with jackknife order indicated by the number of dashes). From top to bottom, we consider extrapolations from 100 to 200, 100 to 400, and 100 to 1000 chromosome. For twofold extrapolation, the third-order weight is a good approximation to the exact weight and the jackknife will be accurate independent of the underlying allele frequency distribution $\Phi(f)$, whereas for 10-fold extrapolation,  the accuracy of the results will depend much more on the cancellation of errors, in the integral of Eq. \eqref{basic}, making results sensitive to model assumptions.  }
\end{figure}

\begin{figure}[h!]
\scalebox{0.55}{\includegraphics{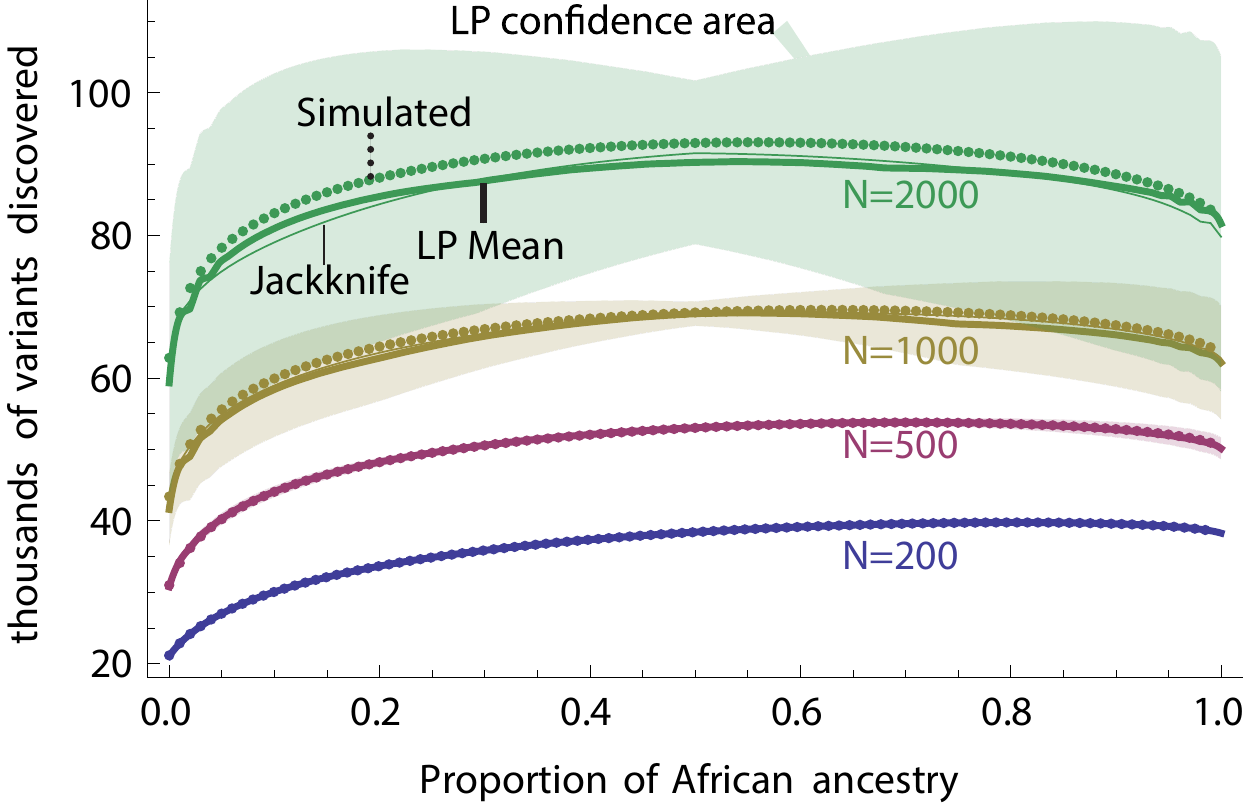}}
\scalebox{0.55}{\includegraphics{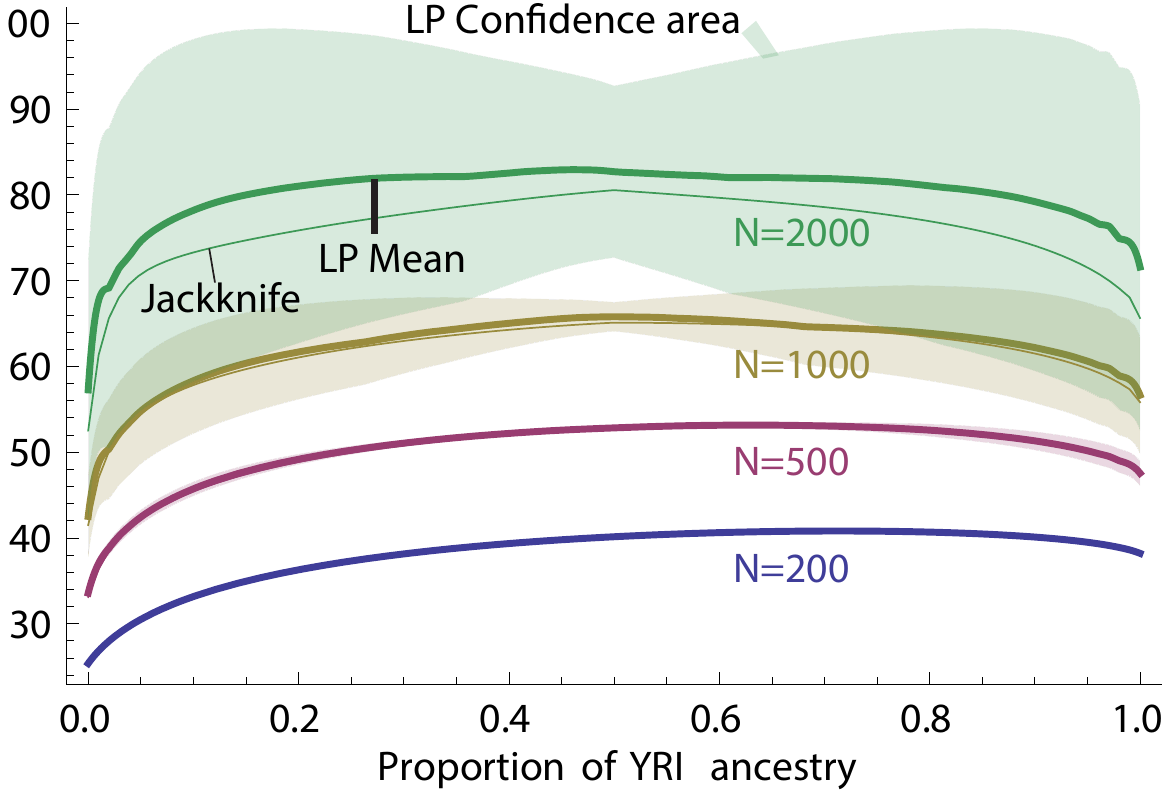}}

\caption{\label{admix} (Left) Predicted and observed discovery rates as a function of sample composition when the sample has both European and West African ancestry, based on a simulated evolutionary model. LP and jackknife predictions for discovery rates were generated using a sample of 100  European and 100 African haplotypes, for varying proportions of European and West African ancestries. These were compared to simulated values according to the model. (Right) Predictions based on 100 haplotypes drawn from 1000 Genomes YRI and CEU samples, as a function of sample composition.}
\end{figure}

\begin{figure}[h!]
\scalebox{1}{\includegraphics{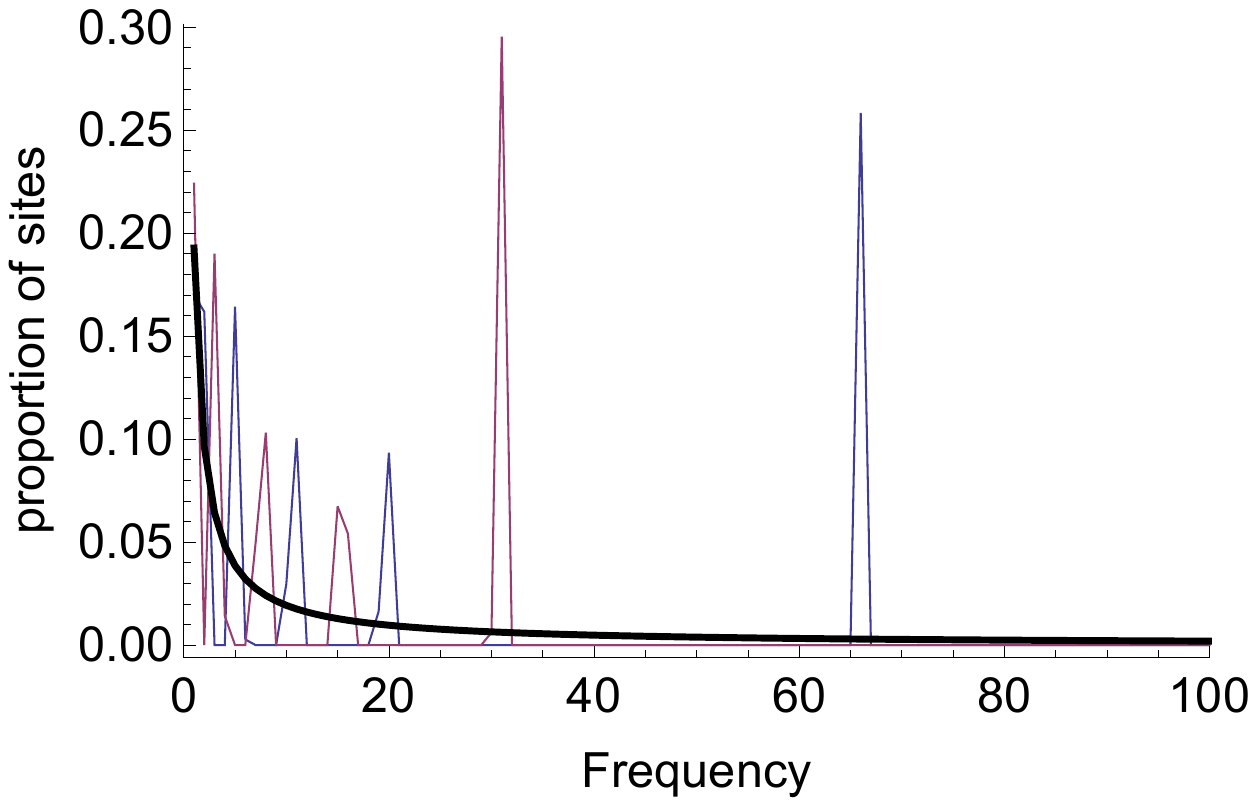}}

\caption{\label{minmax} Three possible SFS' in a sample of size 100 that are consistent with a single simulated observed SFS of size 40.  The black curve is the correct (simulated) SFS in the large sample, and the red (blue) curves were identified by linear programming to provide the maximal (minimal) total number of variants consistent with the data.   Despite the large qualitative differences in the shape of the SFS', the total number of variants differs by less than $1\%$.   }
\end{figure}

\begin{figure}[h!]
\scalebox{1}{\includegraphics{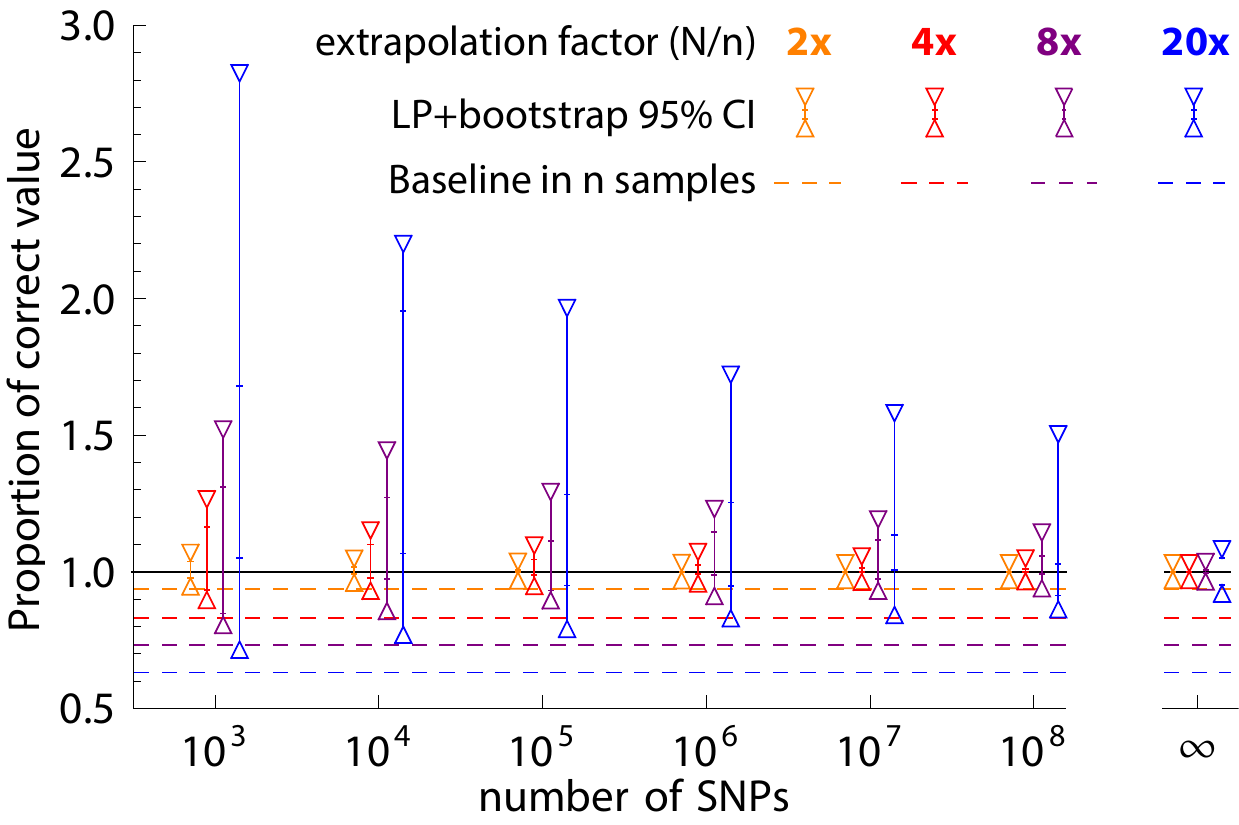}}

\caption{\label{LvNn} The effects of the amount of data on the extrapolation accuracy. We generated Poisson sampling for $10^3$ to $10^8$ polymorphic SNPs in samples of size 100, 200, 400, and 1000. For each, we generated 40 samples of size $50$ by hypergeometric sampling. We obtained upper and lower LP bounds for each simulated set by merging bins until an LP solution is found (see text). Triangle tips represent the upper limit of the $95\%$ CI on the upper bound, and the lower limit of the $95\%$ CI on the lower bound. Vertical lines connect these with the short horizontal lines representing the other end of the respective confidence interval. }
\end{figure}

\end{document}